\documentclass[preprintnumbers,nofootinbib,showpacs,eqsecnum,pre,12pt]{revtex4-1}
\usepackage{amsmath}         
\usepackage{amssymb}          
\usepackage[pdftex]{graphicx}		
\usepackage{color}  				
\usepackage[pdftex]{hyperref} 
\usepackage[utf8]{inputenc}
\usepackage{dsfont}
\usepackage{slashed}

\newcommand{\blue}{\color{blue}}
\newcommand{\green}{\color{green}}
\newcommand{\red}{\color{red}}

\begin{document}

\preprint{XXXX, \today}
\title{Reinterpretation of ATLAS 8 TeV searches for Natural SUSY with a $R$-Sneutrino LSP}
\pacs{XXXX}

 \author{L. Mitzka}
 \email[E-mail: ]{lmitzka@physik.uni-wuerzburg.de}
 \author{W. Porod}
 \email[E-mail: ]{porod@physik.uni-wuerzburg.de}
 \affiliation{
 Institut f\"ur Theoretische Physik und Astropphysik, Uni Wuerzburg
 }

\begin{abstract}
 The data obtained by the LHC collaborations clearly show that supersymmetric models 
 are not realized in nature in a vanilla form and that in particular strongly interacting
 supersymmetric particles are most likely heavier than expected. An exception are the
 partners of the third generation quarks, which also play a dominant role in the breaking
 of the electroweak symmetry.  We consider here an extended class of so-called `natural
 supersymmetric models' where we allow for a sneutrino as the lightest supersymmetric particle
 as it appears for example in left-right symmetric models and/or models where supersymmetry
 is explained via an inverse seesaw mechanism. We evaluate how much existing ATLAS data
 constrain such scenarios and obtain roughly a bound of 300~GeV on the charginos
 if the sneutrinos are lighter than about 120~GeV. For the stop we find that only masses
 up to 300 GeV are excluded independent of the mixing angle. For larger values the exclusion
 depends on the detail of the scenario and if the mass exceeds 800 GeV no bound is obtained.
\end{abstract}

\maketitle

\section{Introduction}

The search for supersymmetry (SUSY) is among the main priorities of the LHC collaborations. Up to now no sign for supersymmetry
or any significant deviation from the Standard Model (SM) prediction has been found. 
In contrast, the
last particle predicted by the SM has been found 
\cite{Aad:2012tfa,Chatrchyan:2012xdj} marking  one of the most important milestones
in particle physics. Its mass is already known rather precisely: 
$m_h = 125.09 \pm 0.21$~(stat.) $\pm 0.11$~(syst.)~GeV \cite{Aad:2015zhl}. Moreover, 
the strengths of the various LHC-signals  are also close
to the SM predictions. The combination of the Higgs discovery with the (yet) unsuccessful searches has led to the
introduction of a model class called `natural  SUSY' 
\cite{Brust:2011tb,Papucci:2011wy,Hall:2011aa,Blum:2012ii,Espinosa:2012in,D'Agnolo:2012mj,%
Baer:2012uy,Younkin:2012ui,Kribs:2013lua,Hardy:2013ywa,Kowalska:2013ica,Han:2013kga}. 
The basic idea of this class of models is to take only those
SUSY particles close the electroweak scale which do give a sizeable contribution to the 
mass of the Higgs boson in order to avoid 
a too large tuning of parameters and to take all other particles at the multi-TeV scale. In particular, the higgsinos (the partners of the Higgs bosons),
the light stop (the partner of the top-quark)  and in case that it is mainly a left-stop also the light sbottom are assigned masses of the order of a
few hundred GeV. In addition the gluino and the heavier stop should be close to the TeV scale.

In the minimal supersymmetric standard model (MSSM) the mass of the Higgs boson is bounded to be 
below the mass
of the $Z$-boson at tree level implying the need of huge radiative corrections close to 90\%
as $m_h^2 \simeq m^2_Z + 86^2$~GeV. Due to the large Yukawa couplings of the top-quark such 
corrections can indeed be
achieved by the requirement of either a large geometric average of the stop masses and/or a large left-right 
mixing parameter $A_t$. 
In non-minimal extensions, the tree-level bound can be pushed to larger values due to the extra 
$F$-contributions
as in the NMSSM 
\cite{Nilles:1982dy,Derendinger:1983bz,Ellis:1988er,Drees:1988fc,Ellwanger:1993xa,%
King:1995vk,Franke:1995tc,Ellwanger:2006rm} or due to extra $D$-term contributions in models 
with an enlarged gauge group
\cite{Haber:1986gz,Drees:1987tp,Cvetic:1997ky,Zhang:2008jm,Ma:2011ea,Hirsch:2011hg}.

The searches for the stop in various channels
have been carried out in the context of the MSSM
by the ATLAS collaboration, see e.g.~\cite{Aad:2013ija,Aad:2014vma,Aad:2014nra,Aad:2015pfx} 
and the CMS collaboration,
see e.g.~\cite{Chatrchyan:2013xna,Khachatryan:2015wza,Khachatryan:2016pup,Khachatryan:2016oia}. 
The reinterpretation of the corresponding data in terms of natural SUSY in the
MSSM context has been carried out in 
\cite{Brooijmans:2014eja,Drees:2015aeo,Kobakhidze:2015scd,Fan:2015mxp}, 
of scenarios based on a compressed spectrum between higgsino and gluino 
in \cite{Chalons:2015vja}, the natural NMSSM in \cite{Kim:2015dpa}
and in a scenario with $R$-sneutrinos in \cite{Guo:2013asa}.

In this paper we will focus on models which can emerge
as a low energy limit from the breaking of an extended gauge sector. 
If these models are left-right symmetric
then one has also to include right-handed neutrinos and their superpartners, the $R$-sneutrinos. 
Here it might well be that actually the lightest
R-sneutrino is the lightest supersymmetric particle (LSP). In these models, the neutrino Yukawa couplings
are rather small if neutrino physics is explained via a low scale seesaw mechanism of type I
 \cite{Minkowski:1977sc,Yanagida:1979as,Mohapatra:1979ia,GellMann:1980vs,Schechter:1980gr}. 
Provided that this results in rather long life-times of the next to lightest SUSY particle (NLSP), e.g.\ in case that the 
stop is the NLSP one expects it to live so long that it can actually leave the detector
\cite{deGouvea:2006wd}. In such cases the searches for $R$-hadrons apply resulting in a mass bound of about $\simeq 900$~GeV \cite{Chatrchyan:2013oca,ATLAS:2014fka}.
Similarly also the case of a chargino NLSP would lead to a stable particle at the relevant time scales of collider measurements. 
However, it might well be that an inverse seesaw mechanism \cite{Mohapatra:1986bd} is the correct explanation of the observed
neutrino data. In this case the neutrino Yukawa couplings are in general large enough that the NLSP decays inside the collider.
Here we will focus on such scenarios and study to which extent the bounds on the stop masses are changed
compared to the MSSM searches. Similar scenarios, motivated by dark matter arguments, 
have been considered in 
\cite{Arina:2015uea,Belanger:2015cra}. This work differs in several aspects: (i) in their
case the resulting R-sneutrino is mainly of  $\tau$ flavour type whereas we will investigate
the case where all three lepton flavours contribute significantly. Also the part of the
parameter space considered differs from ours.
 (ii) They use the package
{\tt SModelS} \cite{Kraml:2013mwa} to obtain the constraints whereas we will use the 
{\tt CheckMATE} \cite{Drees:2013wra}. 
While \texttt{CheckMATE}, \texttt{ATOM} \cite{Mahbubani:2012qq,Papucci:2014rja}, \texttt{PGS} \cite{PGS} or \texttt{MadAnalysis} \cite{Conte:2012fm,Conte:2014zja,Dumont:2014tja} 
deduce limits with a detector simulation for generated events, \texttt{SModelS} or \texttt{Fastlim} \cite{Papucci:2014rja} 
constrain models by comparing the cross-section times branching ratio for certain topologies with the observed upper limits 
taking the respective efficiencies into account. \texttt{PGS} has been
used in \cite{Guo:2013asa} and there 
only the case of a pure right-handed stop $\tilde t_R$ has been considered in detail, 
whereas we will present
here results for various stop mixing angles. 

In the next section we will introduce the model focussing on the parts relevant for this investigation.
In section \ref{sec:scan} we summarize the relevant LHC analyses as implemented in the package \texttt{CheckMATE}
\cite{Drees:2013wra}, as well as the generation of the underlying Monte-Carlo data. In section \ref{sec:results}
we present our results and conclude finally in section \ref{sec:conclusion}.

\section{The Model}

The simplified model under consideration is based on the relevant Natural SUSY particle content, 
where one adds to the SM the superpartners of the third generation squarks, the two
stops $\tilde t_{1,2}$ and the two sbottoms
 $\tilde b_{1,2}$, and the partners of the
Higgs-bosons, the so-called higgsinos consisting of two neutral states and a charged one.
As usual we will assume that the heavier sbottom $\tilde b_{2}$ is too heavy to contribute
wherever possible but it will turn out that in some case it has to be included for consistency.
Moreover,
 we add three $R$-sneutrinos $\tilde \nu_{R,k}$ ($k=e,\mu,\tau$) which take
over the role of the lightest supersymmetric particle(s) (LSP) and, thus, they can form the
dark matter of the Universe as we assume conserved $R$-parity,
see e.g.~\cite{Gopalakrishna:2006kr,Asaka:2006fs,Lee:2007mt,Arina:2007tm,Thomas:2007bu,%
Deppisch:2008bp,Cerdeno:2009dv,Kumar:2009sf,Belanger:2011rs,DeRomeri:2012qd}. 
To be more precise,
we consider a scenario where neutrino-data are explained via an inverse seesaw and, thus,
one needs in addition three fields $\tilde S_k$ which carry lepton number as well. 
However,
for our purposes they are essentially decoupled apart from the mixing with the $R$-sneutrinos.
The same holds for the heavy neutrino-states. In principle they could show up in the
decays of the SUSY particles but the corresponding decay widths are kinematically suppressed.
All other supersymmetric particles are assumed to be too heavy to play any role here.
The relevant part of the superpotential reads as 
\begin{align}
\mathcal{W}_{eff} = \mu \widehat{H}_u \cdot\widehat{H}_d 
 + Y_t \hat{t}_R \widehat{H}_u\cdot\widehat{Q} + Y_b \widehat{b}_R\widehat{Q} \cdot \widehat{H}_d
  + \sum_k \left(
 Y_{\nu,k} \hat{\nu}_{R,k} \widehat{H}_u 
 \cdot\widehat{L}_k  + M_k \widehat{S}_k \hat{\nu}_{R,k} \right)\quad ,
\end{align}
where the $\cdot$ indicates the $SU(2)$ invariant product. For the explanation of neutrino 
data one would have to add a term $\sum_{jk} \mu_{jk} \widehat{S}_j \widehat{S}_k$
where max($|\mu_{jk}|$) has to be much smaller than any of the masses discussed here.
Therefore, we can safely neglect it here as its effect would be a tiny mass splitting
between the real and imaginary parts of the sneutrinos, see e.g.~\cite{Hirsch:2012kv}, which is not
measurable at the LHC or even an ILC. The existence of the $\mu_{jk}$ allows us also
to take $Y_{\nu}$ as flavour diagonal.
The corresponding soft SUSY breaking terms are given by
\begin{align}\mathcal{V}^{soft} =&\frac{1}{2}M_3 \tilde{g}\tilde{g} +
   m^2_{H_d} \left|H_d\right|^2 
   +  m^2_{Q} \left|Q\right|^2 
   + m^2_{b_R} \left|b_R\right|^2 + m^2_{t_R} \left| t_R \right|^2
  + m^2_{\nu_R} \left| \tilde{\nu}_R \right|^2 \nonumber \\
 & + B_\mu H_u \cdot H_d +  \sum_k B_{M_k} \tilde{S}_k \tilde{\nu}_{R,k}
 + T_t \tilde{t}_R H_u \cdot  \tilde{Q}+ T_b \tilde{b}_R \tilde{Q} \cdot H_d 
  + T_\nu \tilde{\nu}_R \tilde{H}_u\cdot \tilde{L} \quad .
\end{align}

It is well known that there can be a sizeable mixing between L- and R-states in case of
third generation squarks which is expressed by
\begin{align}
\begin{pmatrix}
\tilde{t}_1 \\ \tilde{t}_2
\end{pmatrix}
= 
\begin{pmatrix}
\cos\theta_{\tilde{t}} & -\sin\theta_{\tilde{t}} \\
\sin\theta_{\tilde{t}} & \cos\theta_{\tilde{t}}
\end{pmatrix}
\begin{pmatrix}
t_L \\ t_R
\end{pmatrix}
\quad ,\quad
\begin{pmatrix}
\tilde{b}_1 \\ \tilde{b}_2
\end{pmatrix}
= 
\begin{pmatrix}
\cos\theta_{\tilde{b}} & -\sin\theta_{\tilde{b}} \\
\sin\theta_{\tilde{b}} & \cos\theta_{\tilde{b}}
\end{pmatrix}
\begin{pmatrix}
t_L \\ t_R
\end{pmatrix}\quad .
\end{align}
Here we use the convention that $m_{\tilde t_1}\le m_{\tilde t_2}$ and 
$m_{\tilde b_1}\le m_{\tilde b_2}$.
In the following we will take the masses of the physical states as well as the mixing angles
$\theta_{\tilde{t}}$ and $\theta_{\tilde{b}}$ as free parameters. Note that only five out of
the six parameters can be chosen freely as the parameter $m^2_Q$ appears in the mass matrices
of both, $\tilde t$ and $\tilde b$ implying the following relation
\begin{align}
  m_W^2\cos 2\beta \, = \, m_{\tilde t_1}^2\cos^2\theta_{\tilde t} + m_{\tilde t_2}^2\sin^2\theta_{\tilde t}
  - m_{\tilde b_1}^2\cos^2\theta_{\tilde b} - m_{\tilde b_2}^2\sin^2\theta_{\tilde b} - m_t^2 + m_b^2
\label{eq:squarkmassrelation}
\end{align}
at tree-level. Therefore we take as input $m_{\tilde t_1}$, $m_{\tilde t_2}$, $m_{\tilde b_1}$,
$\theta_{\tilde{t}}$ and $\theta_{\tilde{b}}$. In case of $\theta_{\tilde{t}}=0$ it
can happen that the calculated $m_{\tilde b_2}$ is actually smaller than the input value
for $m_{\tilde b_1}$. In such cases we will relabel the corresponding states according to
the correct mass ordering.

In the slepton sector we have a mixing between $\tilde \nu_R$ and $\tilde S$. Neglecting
generation mixing one finds
the following mass matrix
\begin{equation}
m^2 = \left( \begin{array}{cc}
|M_k|^2 & B_{M_k} \\
B_{M_k} & |M_k|^2 \end{array} \right)  \,.
\end{equation} 
For completeness we note that there are corrections proportional to the left-right mixing
in the sneutrino sector of the L-slepton masses squared which we neglect here as the L-sleptons
are assumed to be very heavy. Provided that the $|M_k|^2$ has a similar size as $B_{M_k}$
one finds a rather light state with mass of order $|M_k|^2-|B_{M_k}|$ 
and a rather heavy state with mass of $|M_k|^2+|B_{M_k}|$ and a nearly maximal mixing.
Despite the nearly maximal mixing we still call the light states $\tilde \nu_R$ in an obvious
abuse of language. 
We will neglect for simplicity in the following generation mixing in this 
sector
 and
assume that the three lightest states are mass-degenerate to reduce the number of free
parameters. In the same spirit we will assume that $Y_\nu$ is diagonal and that all
entries have the same size. This set-up is equivalent to assuming only one
light $\tilde \nu_R$-state which couples with equal strength to all charged leptons
and the charged Higgs boson. We will also comment on the expected changes when departing
from these assumptions.  

The masses of the higgsino-like neutralinos $\tilde{\chi}^0_{1/2}$ and the chargino 
$\tilde{\chi}^\pm_{1}$ are given by $|\mu|$ up to small corrections 
of order ${\cal O}(m_Z/$min$(|M_1|,|M_2|) )$, see e.g.\ \cite{Barducci:2015ffa}.
The resulting mass differences are at most a few GeV and will be neglected in the following
because the visible products of the corresponding decays, e.g.\ 
$\tilde \chi^+_1 \to \tilde \chi^0_1 \pi^+$, are very soft and cannot be detected at the LHC.
Thus we will set $m_{\tilde{\chi}^\pm_1} = m_{\tilde{\chi}^0_{1/2}}=|\mu|$. Last but not
least we assume also the gluino to be too heavy to be produced at the LHC with 
$\sqrt{s}=8$~GeV at a sizeable rate.
Summarizing, we have thus the following set of eight free parameters: 
$m_{\tilde t_1}$, $m_{\tilde t_2}$, $m_{\tilde b_1}$,
$\theta_{\tilde{t}}$, $\theta_{\tilde{b}}$, $m_{\tilde \nu_R}$, $Y_\nu$ and $\mu$.

We assume, as mentioned above, that the $\tilde \nu_R$ is the LSP and, thus, we
have essentially two different mass orderings depending on whether the squarks are
lighter or heavier than the higgsinos.  In the second case the squarks decay according
to 
\begin{eqnarray}
\label{eq:squark-decays}
\tilde t_i &\to & t \tilde{\chi}^0_{1/2} \,\,,\,\, b \tilde{\chi}^+_1 \\
\tilde b_i &\to & b \tilde{\chi}^0_{1/2} \,\,,\,\, t \tilde{\chi}^-_1 
\end{eqnarray}
with $i=1,2$. If there is a sufficient large mass splitting, then there are in addition 
bosonic final states \cite{Bartl:1994bu} which we take into account
as well
\begin{align}
\label{eq:pure-bosonic-processes}
\tilde{t}_1 \to W^+ \tilde{b}_i \,,  \quad \tilde{b}_i &\to W^- \tilde{t}_1     \\
\tilde{t}_2 \to \tilde{t}_1 h^0_1  \,,  \quad \tilde{t}_2 &\to \tilde{t}_1 Z^0 \,, \quad     
\tilde{b}_2 \to \tilde{b}_1 h^0_1 \,,   \quad \tilde{b}_2 \to \tilde{b}_1 Z^0     
\end{align}
whose relative importance depends on the mixings in the stop- and sbottom sectors, respectively.
In case that only the fermionic decay modes are allowed at a sizable rate, then
the final states of stops and sbottoms are practically the same in the standard Natural
SUSY scenarios as the decay products of
of $\tilde{\chi}^0_{2}$ and $\tilde{\chi}^\pm_1$ are too soft to be detected, e.g.\
all higgsino like states lead to the same missing energy signatures as the LSP.
In our scenario the higgsinos will decay further to 
\begin{align}
\label{eq:chargino-decay}
\tilde{\chi}^+_1     &\to \tilde{\nu}_{Ri} \ell^+_i \\
\label{eq:neutralino-decay}
\tilde{\chi}^0_{1/2} &\to \tilde{\nu}_{Ri} \nu_i
\end{align}
with $\ell_i = e,\mu,\tau$. This results in cascade decays with final states like
\begin{align}
\label{eq:threebody-decays}
\tilde{t}_1 \to b l^+ \tilde{\nu}_{Ri} \quad , \quad &\tilde{t}_1   \to t \nu_i \tilde{\nu}_{Ri}
 \quad  ,\\
\tilde{b}_1 \to t l^- \tilde{\nu}_{Ri} \quad , \quad &\tilde{b}_1 \to b \nu_i \tilde{\nu}_{Ri}
 \quad .
\end{align}
In case that the higgsinos are heavier than the squarks one has the same final states which
are now mediated via off-shell higgsinos\footnote{In our calculation we take also the off-shell
gauginos into account in this case whose contribution is however suppressed as they are
substantially heavier than the higgsinos.} resulting in somewhat changed angular distributions
between the charged fermions, e.g.\ in the $b\,l$ and the $t\,l$ systems.
\begin{figure}[t]
\includegraphics[scale=0.6]{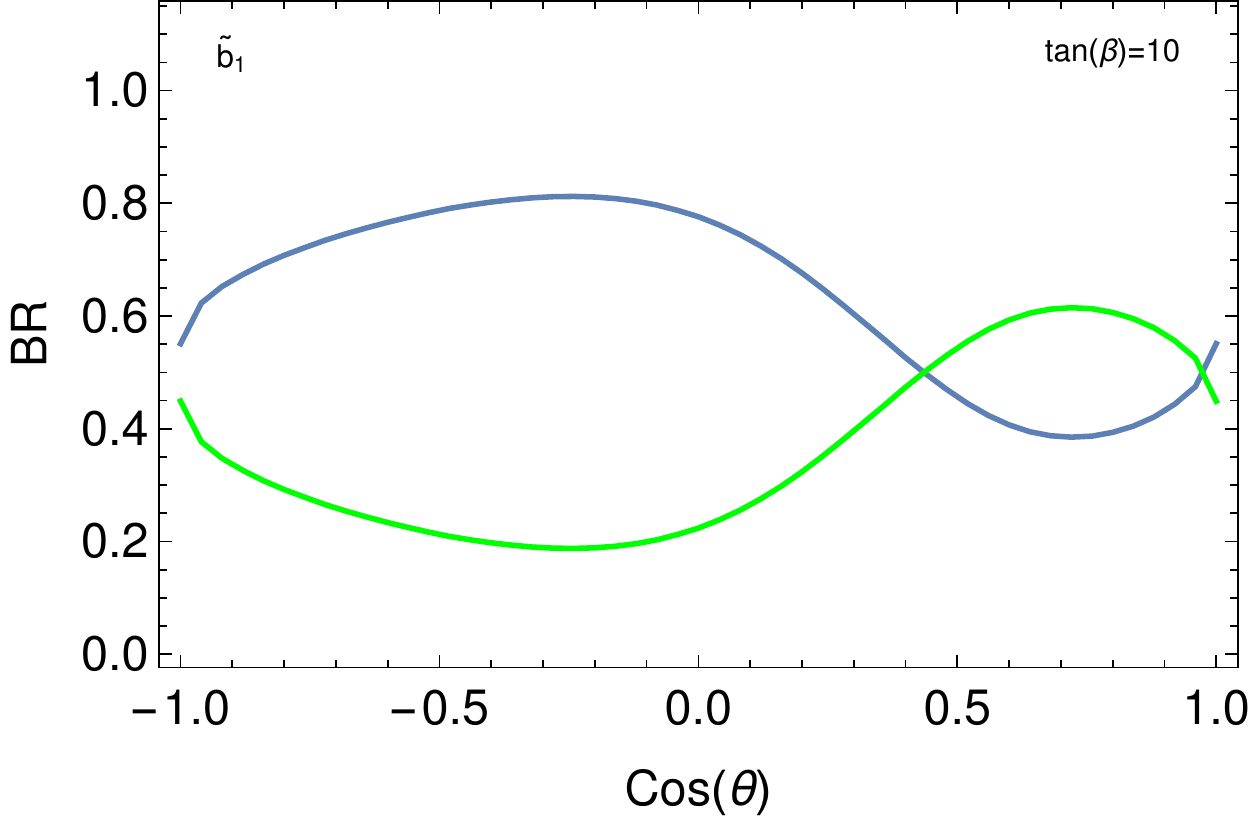}
\includegraphics[scale=0.6]{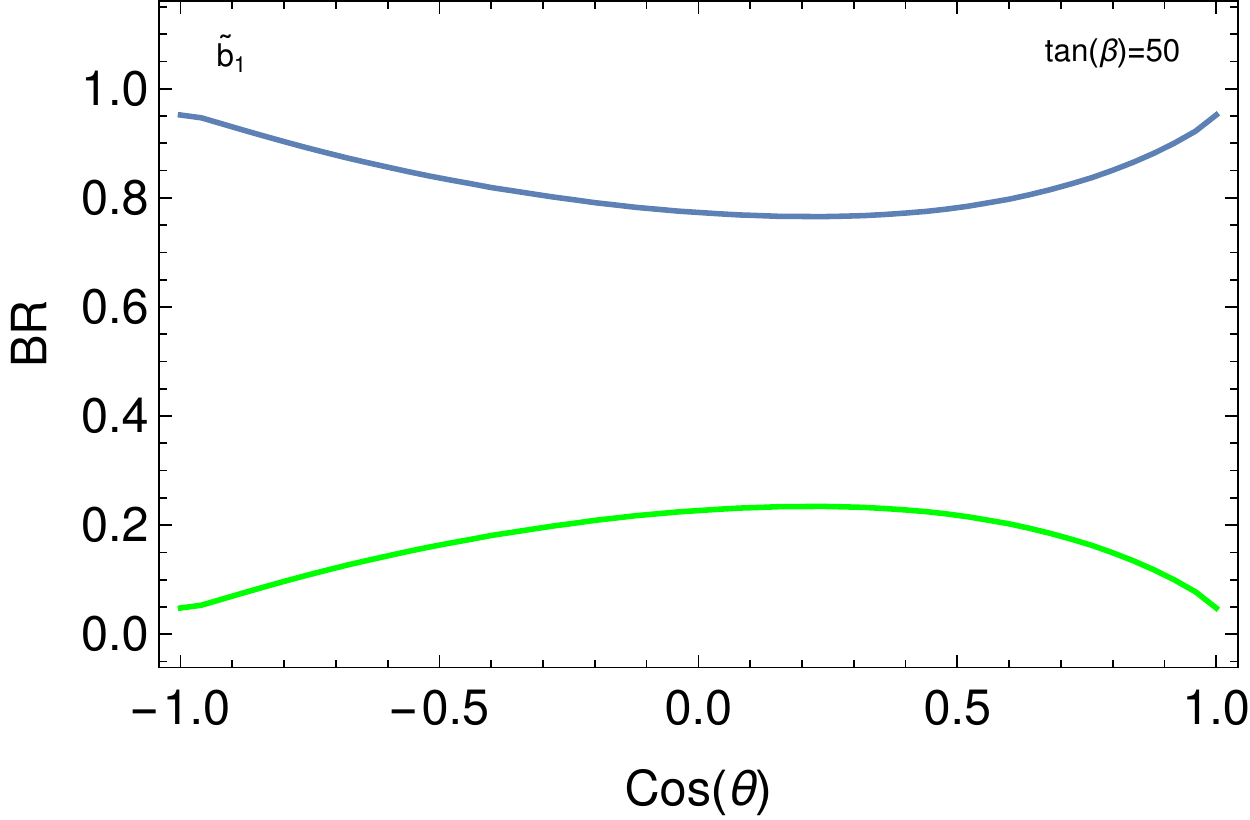} \\[4mm]
\includegraphics[scale=0.6]{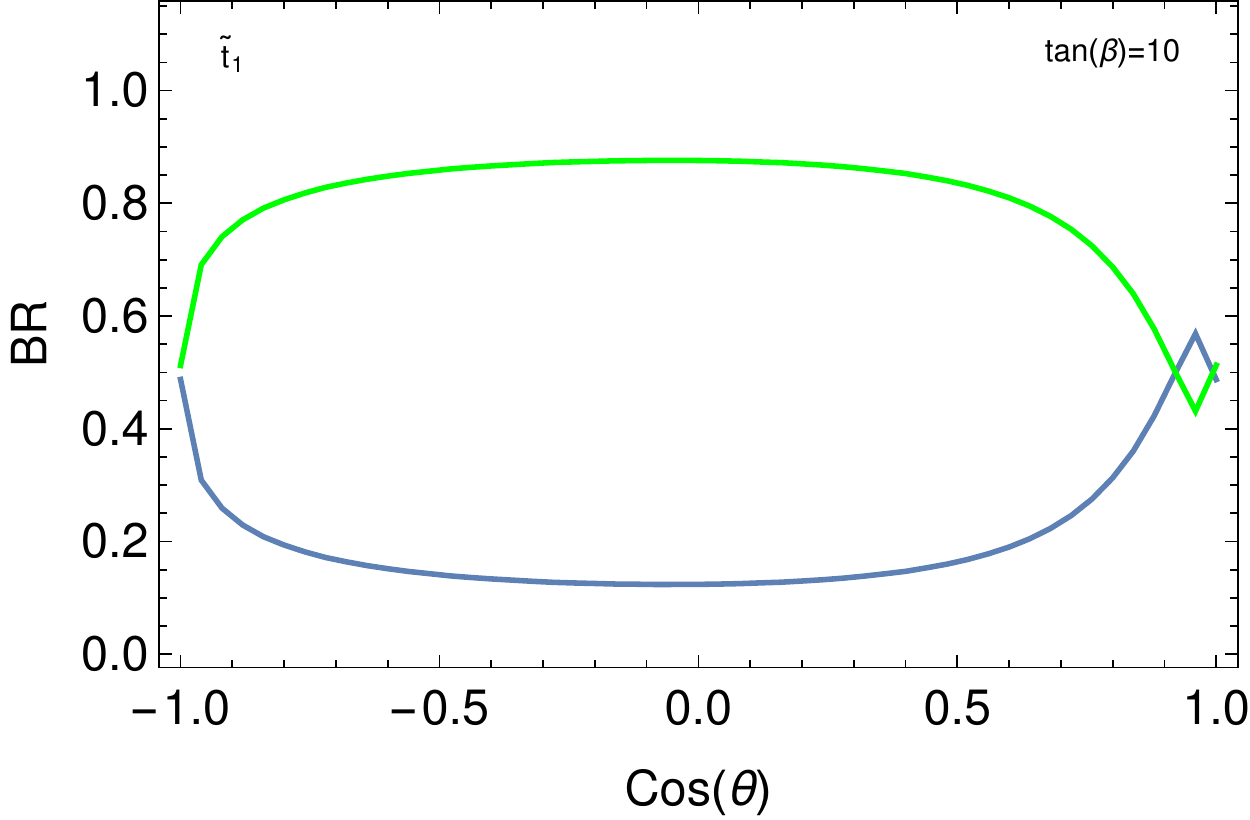}
\includegraphics[scale=0.6]{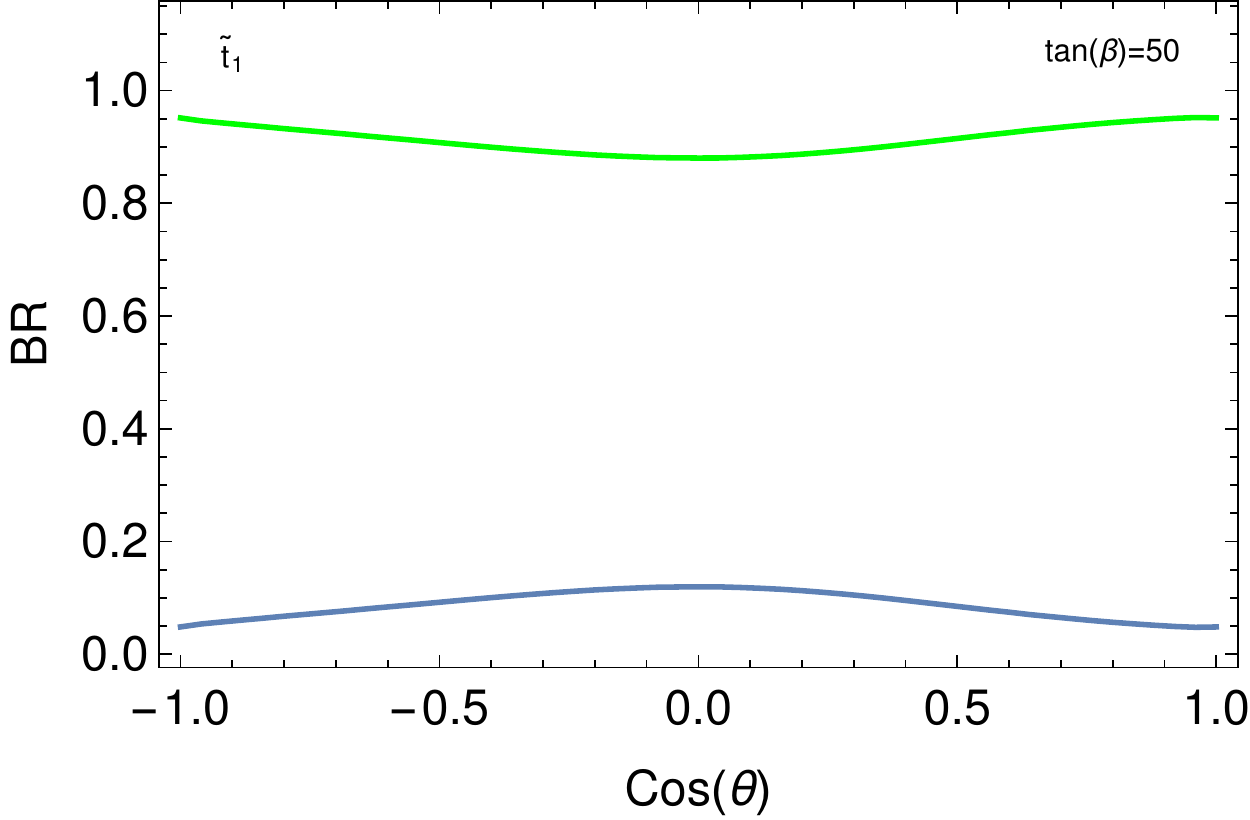} 
\caption{Branching ratios of the light stop and the light sbottom taking 
$m_{\tilde q_1}=500$~GeV ($q=b,t$), 
 $m_{\tilde \nu_R}=100$~GeV, $\mu=590$~GeV, $M_1=M_2=1$~TeV. The upper row shows
sbottom decays, the lower row stop decays. The left column displays the case $\tan\beta=10$ and 
right one $\tan\beta=50$. The lines correspond to $\tilde q_1 \to  q \nu \tilde \nu_R$ (blue line)
and $\tilde q_1 \to  q' l \tilde \nu_R$ (green line) summing over all lepton flavours. By construction
each lepton flavour has the same probability. 
}
\label{fig:brs}
\end{figure}
A typical example for the branching ratios is shown in Fig.~\ref{fig:brs}. An important observation is that for
large $\tan\beta$ the final states containing a $t$-quark get reduced and correspondingly
the final states with a $b$-quark are enhanced. The reason is that for smaller $\tan\beta$ the
kinematical differences are compensated to some extent by the differences in the corresponding
Yukawa couplings. The kink in the stop branching ratios close to $\cos \theta_{\tilde t}=0.95$
is due to a negative interference between the higgsino component, which is suppressed by
$\sin \theta_{\tilde t}$, and the gaugino component of the 
$\tilde t_1$-$b$-$\tilde \chi^+_1$  coupling.

\section{Set-up and parameter space scan}
\label{sec:scan}

For this investigation we have used a series of public programs: 
As a first step we have used \texttt{SARAH} 
\cite{Staub:2008uz,Staub:2013tta,Staub:2012pb,Staub:2010jh,Staub:2009bi} in the 
\texttt{SUSY/BSM toolbox 1.2.9} \cite{Staub:2011dp,Staub:2015kfa} to implement 
the aforementioned model into the event generator  \texttt{WHIZARD 2.2.6} 
\cite{Kilian:2007gr,Moretti:2001zz}.
We use the \texttt{CTEQ6L1} PDF set \cite{Pumplin:2002vw} in \texttt{WHIZARD}, which uses 
\texttt{PYTHIA 6.427} \cite{Sjostrand:2006za} 
internally for showering and hadronization with the ATLAS 
\texttt{AUET2B-CT6L} tune to generate events at tree-level. 
The parameter scan has been automated using \texttt{gnu-parallel} \cite{Tange2011a}.
As default we generate  25000 events for every production process via strong interactions, 
e.g.\ for $pp\to \tilde t_1 \tilde t_1^*$. 
The generated events are then fed in the \texttt{HEPMC} \cite{Dobbs:2001ck} format into \texttt{CheckMATE 1.2.0} \cite{Drees:2013wra} that uses 
internally a modified version of \texttt{Delphes} as detector simulation \cite{deFavereau:2013fsa},
\texttt{FastJet} \cite{Cacciari:2011ma,Cacciari:2005hq} to define jets via the anti-$k_T$ algorithm 
\cite{Cacciari:2008gp}. 
These event samples are then re-weighted according to the corresponding total
 cross-section which we compute with \texttt{PROSPINO 2.1}
  \cite{Beenakker:1996ch,Beenakker:1997ut} to get a more reliable result.
\texttt{CheckMATE} compares the number of events passing each signal region 
of every considered analysis with the observed $S95$ limit obtained by the ATLAS collaboration via 
the parameter
\begin{align}
\label{eq:r-value-definition}
r^c_{exp/obs} = \frac{S-1.96\cdot\Delta S}{S^{95}_{exp/obs}}
\end{align}
with $S$ being the number of the considered signal region, $\Delta S$ as the error from the Monte 
Carlo and $S^{95}_{exp/obs}$ is the expected or experimentally observed 95\% confidence limit on the signal 
\cite{Drees:2013wra,Drees:2015aeo}. 

\texttt{CheckMATE} contains also CMS analyses which we
did not exploit here  for a practical reason: 
in the version used one can combine easily different analyses carried out by one collaboration
in a single run but not those carried out by two different collaborations. Thus their
inclusion would have nearly doubled our computation time. In view of the fact, that both
collaborations have obtained rather similar results, we do not expect any significant
differences from the results obtained here compared to the combined analyses of both
collaborations. The analyses used here are listed in table~\ref{tab:ListOfAnalyses}
together with their main characteristics.

\begin{table}[t]
\caption{List of the ATLAS analyses used in this study.}
\label{tab:ListOfAnalyses}
\begin{center}
\begin{tabular}{c c c c}
\hline 
Ref. & \texttt{CheckMATE} analysis name & Search for\dots & \dots in finals states with\dots\\ 
\hline
\textbf{multilepton:}  &  &  & \\ 
\cite{Aad:2014pda} & \texttt{atlas\_1403\_2500} &  $\tilde{g}$ and $\tilde{q}$ &  jets, 2SS/3 leptons \\ 
\cite{ATLAS-CONF-2013-036} & \texttt{atlas\_conf\_2013\_036} & $R$PV \& $R$PC SUSY &  four or more leptons \\ 
\cite{Aad:2014nua} & \texttt{atlas\_1402\_7029} & $\tilde{\chi}^\pm$ and $\tilde{\chi}^0$ &  3 leptons and $\slashed{E}_T$ \\
\hline
\textbf{dilepton:} &  & &  \\
 
\cite{Aad:2014qaa} & \texttt{atlas\_1403\_4853} & $\tilde{t}$ & two leptons and 2 $b$ jets  \\ 
\cite{Aad:2014vma} & \texttt{atlas\_1403\_5294} & $\tilde{\ell},\tilde{\chi}^{0,\pm}$ &  two leptons and $\slashed{E}_T$ \\
\cite{ATLAS-CONF-2013-089} & \texttt{atlas\_conf\_089} & $\tilde{t}$ &  two leptons via the razor variable \\
\cite{ATLAS-CONF-2013-049} & \texttt{atlas\_conf\_2013\_049} & $\tilde{\chi}^{0,\pm},\tilde{\ell}$ &  two leptons\\
\cite{ATLAS-CONF-2014-014} & \texttt{atlas\_conf\_2013\_014} & $\tilde{t}$ & 2 $b$ jets, two leptons (vie $\tau$), $\slashed{E}_T$ \\

\hline
\textbf{single lepton:} &  &  & \\ 

\cite{Aad:2014kra} & \texttt{atlas\_1407\_0583} & $\tilde{t}$ & 1 lepton, jets and $\slashed{E}_T$  \\ 
\cite{ATLAS-CONF-2013-062} & \texttt{atlas\_conf\_2013\_062} & $\tilde{t},\tilde{g}$ & 1 lepton, jets and $\slashed{E}_T$ \\
\cite{ATLAS-CONF-2012-104} & \texttt{atlas\_conf\_2013\_104} & $\tilde{t}$ & 1 lepton, jets and $\slashed{E}_T$ \\

\hline
\textbf{hadronic:} &  & & \\ 

\cite{ATLAS-CONF-2013-061} & \texttt{atlas\_conf\_2013\_061} & $\tilde{g}$ &  three $b$-jets and $\slashed{E}_T$ \\ 
\cite{Aad:2013ija} &\texttt{atlas\_1308\_2631}  & $\tilde{b},\tilde{t}$ &  2 $b$ jets and $\slashed{E}_T$ \\ 
\cite{ATLAS-CONF-2013-047} & \texttt{atlas\_conf\_2013\_047} & $\tilde{q},\tilde{g}$ &  jets and $\slashed{E}_T$ \\
\cite{ATLAS-CONF-2013-024} & \texttt{atlas\_conf\_2013\_024} & $\tilde{t}$ & hadronic $t\bar{t}$ final states \\ 
 &  & & \\  
  &  & & \\ 
\end{tabular} 
\end{center}
\end{table}

The fact, that we consider a $\tilde \nu_R$~LSP implies that we expect on average
more charged leptons in the final state than in the conventional Natural SUSY
scenarios. This implies that the relative importance of the analyses gets changed on which we
want to comment. Obviously the realm of multilepton searches is of much bigger importance 
in our case, in particular due to the better detectability of the leptons 
compared to jets. A prominent example where this is important are those containing
the decay $\tilde{b}_1\to\tilde{\chi}^\pm t \to l\tilde{\nu}_R t$, where
the chargino can be either on- or off-shell, and additional lepton(s) from the $t$  decay(s). 
Another example are scenarios containing decays like 
$\tilde{t}_1\to\tilde{b}W^\pm\to t\ell\tilde{\nu}_R W^\pm$ where the $W$ decays into leptons.

Dilepton searches retain their importance, especially in the derivation of limits for $\tilde{t}_1$ 
via the process $\tilde{t}\to\tilde{\chi}^\pm b\ell$ via an on- or off-shell $\tilde{\chi}^\pm_1$ in 
the $2\ell 2b \slashed{E}_T$ final state. 
This has also been observed in \cite{Guo:2013asa}
where exclusion limits considering only $\tilde{t}$ via the 
process $\tilde{t}_1\to \tilde{\nu}_R b\ell$ have been investigated in.
Monolepton searches, mostly in combination with the requirement of two tagged $b$-jets are 
also of significance in cases where the above mentioned scenario either competes with a 
'standard' decay $\tilde{b}_1\to b \tilde{\chi}^0_{1/2}\to b\nu\tilde{\nu}_R$, or the 
process $\tilde{t}_1 \to t\nu\tilde{\nu}_R$ is the dominant decay channel for the
 $\tilde{t}_1$, such one arrives at the well known $t+\slashed{E}_T$ signal.

There are also scenarios where purely hadronic searches are important.
From these especially \texttt{atlas\_1308\_2631}, where a search for $\tilde{t},\tilde{b}$ via 
$2b+\slashed{E}_T$ is presented, is important as it covers the case of 
$\tilde{b}\to b\nu\tilde{\nu}_R$ and exploits the presence of tagable $b$-jets in 
the corresponding final states. Note, however, that this analysis relies on requirements of very 
hard jets, where the leading jet is required to have $p_T>150$ GeV, and large missing energy.
Here one has potentially only a small acceptance in cases with moderate or small mass differences. 
This is especially pronounced in its signal region B, that focusses on small mass differences 
between third-generation squark and LSP with the trade-off to need hard ISR or FSR. 
As we only simulate the tree-level process without hard initial state radiation (ISR) and/or 
final state radiation (FSR) we often do not meet the 
requirements of this signal region. However, we take the resulting uncertainty into 
account in the interpretation of the {\tt CheckMATE} result.
We follow ref.~\cite{Drees:2015aeo} in the categorization of the \texttt{CheckMATE} results: 
we  define every point as either `strictly allowed' if $r^c_{obs} < \frac{2}{3}$, 
as `strictly excluded'  if $r^c_{obs} > 1.5$ and `inconclusive' or `ambiguous"  in case of
$\frac{2}{3} < r^c_{obs} < 1.5$. This aids in keeping our statements less dependent on statistical 
fluctuations and thus more conservative. Moreover, we do not expect that the inclusion
of higher order corrections will change the $r$ value such, that it switches from 
`strictly allowed' to `strictly excluded' or vice versa.
Our procedure of labelling a parameter space combination as `strictly excluded'  
differs slightly from the \texttt{CheckMATE} procedure:
While \texttt{CheckMATE} takes the $r_{obs}^c$ value of the largest $r_{exp}^c$ to find its 
exclusion statement, we also check all other analyses with $r_{exp}^c > 1.0$ for $r_{obs}^c > 1.5$. 
If there is at least one analysis that excludes this parameter point, we take it as strictly 
excluded. If there is no such analysis, we take the statement given by \texttt{CheckMATE}.

We have chosen a regular grid in the parameter space instead of performing a random
scan to get a better understanding of the different features and potential pitfalls 
when interpreting the results. The grid is given by combining the following parameters for
the values given:
  \begin{itemize}
  \item $m_{\tilde{t}_1}$ in GeV:  300, 400, 500, 600, 700, 800, 900, 1000
  \item $m_{\tilde{b}_1}$ in GeV:  300, 400, 500, 600, 700, 800, 900, 1000 
  \item $m_{\tilde{\nu}_R}$ in GeV : 60, 100, 200, 300, 400, 500
  \item $\mu$ in GeV: 110, 190, 290, 390, 490, 590 and require $m_{\tilde{\nu}_R} < \mu$
  \item $\tan\beta$: 10, 50
  \item $\theta_{\tilde{t}}$: $0^{\circ}, 45^{\circ}, 90^{\circ}$
  \item $\theta_{\tilde{b}}$: $0^{\circ}, 45^{\circ}, 90^{\circ}$
  \item $M_1=M_2=1$~TeV 
   \item everything else, including $\tilde{t}_2$,$\tilde{b}_2$ and $m_{\tilde{g}}$: 2 TeV\\
   The exception is potentially $m_{\tilde{b}_2}$ 
   when eq.~(\ref{eq:squarkmassrelation}) becomes effective as mentioned above. 
 \end{itemize} 

\section{Results}
\label{sec:results}

An obvious constraint for our model are searches for electroweakinos and sleptons which 
probe the final state resulting from the pair production of charginos and their decays:
\begin{align}
\label{eq:electroweak-production}
pp\to \tilde{\chi}^+_1\tilde{\chi}^-_1\to\ell^+\ell^{-}\tilde{\nu}_R\tilde{\nu}^{*}_R \quad .
\end{align}
This decay of pair produced $\tilde{\chi}^\pm_1$ which can already restrict the 
$(m_{\tilde{\chi}} , m_{\tilde{\nu}_R})$ mass plane in the dilepton channel.
However the related process in eq. (\ref{eq:neutralino-decay}) is much more challenging, as it would manifest itself only in a monojet signature \cite{Dreiner:2012gx}. It has
been shown in \cite{Barducci:2015ffa} that existing LHC data does not restrict the values for
$\mu$ considered here from mono-jet searches and, thus, our findings are not affected by
the fact that we do not take ISR into account here.

\begin{figure}[t]
\begin{center}
\includegraphics[scale=0.5]{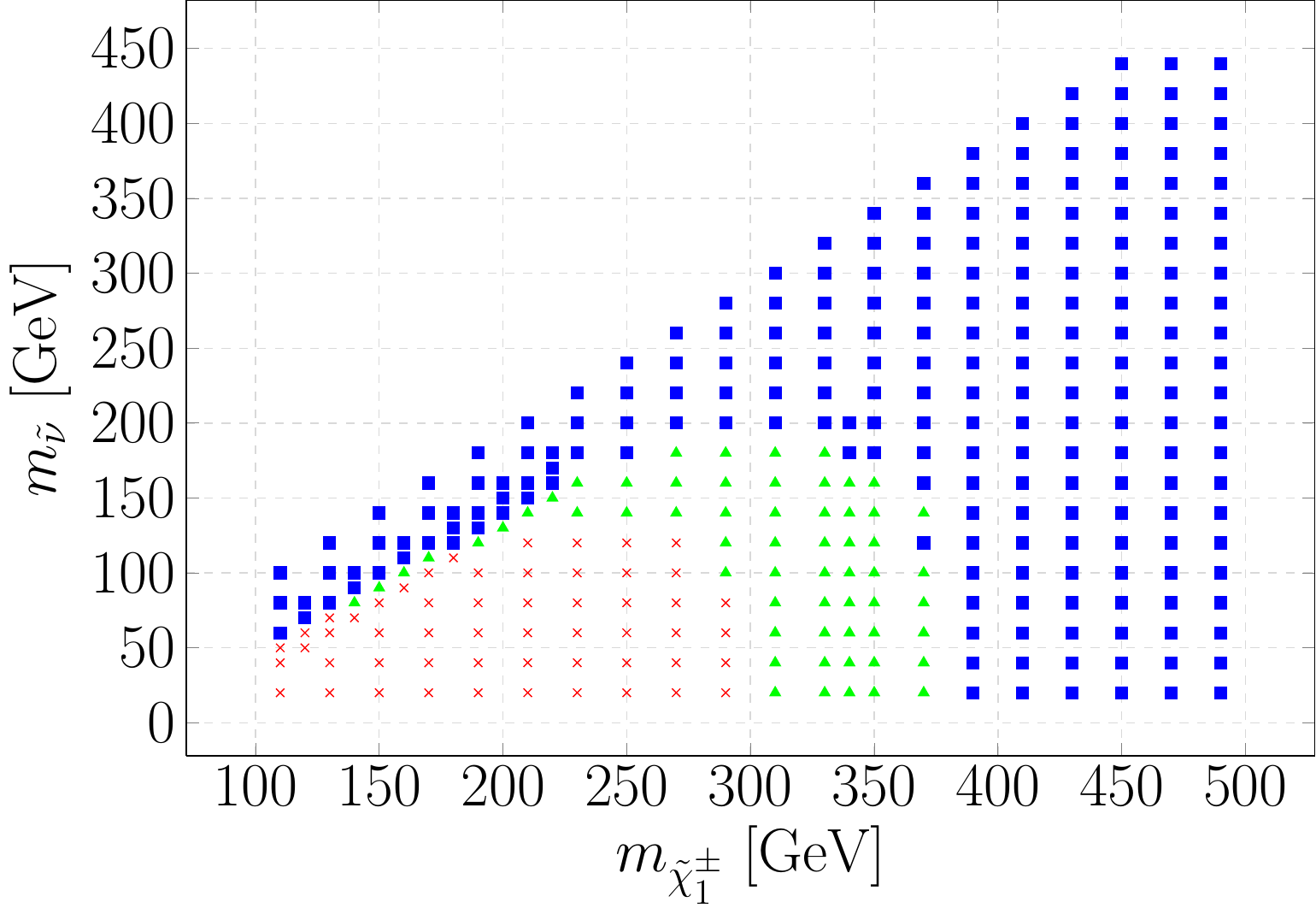}
\end{center}
\caption{Exclusions in the $(m_{\tilde{\chi}^\pm},m_{\tilde{\nu}_R})$ mass plane derived from the 
ATLAS searches \cite{ATLAS-CONF-2013-049} and  \cite{Aad:2014vma}:
points with a {$\red \times$} are excluded,
{$\green \blacktriangle$} designate points as ambigous,
 whereas  
points with {$\blue\blacksquare$} are still allowed.}
\label{fig:chargino-sneutrino-exclusion}
\end{figure}

From the relevant searches, \texttt{atlas\_conf\_2013\_049}  \cite{ATLAS-CONF-2013-049} 
and \texttt{atlas\_1403\_5294} \cite{Aad:2014vma},
the latter is more important than the first one as already noted in  \cite{Arina:2015uea}.
The reason is that the decay kinematics of $\tilde{\ell}^\pm\to\ell\tilde{\chi}^0_1$ is
essentially the same as in our case whereas those of 
$\tilde \chi^+_1 \to \tilde{\chi}^0_1 W^+$ show a substantial difference.
The impact of these analyses on the parameter space is shown in
 Fig.~(\ref{fig:chargino-sneutrino-exclusion}).
This allows to exclude parameter combinations containing the 
$(m_{\tilde{\chi}^\pm},m_{\tilde{\nu}_R})$ combinations
\mbox{$(190,60),(190,100)$} and $(290,60)$ from further consideration.

For the investigation of the remaining parameter space we calculate as default the rates
for the processes 
\begin{align}
\label{eq:third-gen-production1}
pp\to \quad\tilde{b}_1\tilde{b}_1^*,\quad \tilde{t}_1\tilde{t}_1^*\quad .
\end{align}
For scenarios where the application of eq.~(\ref{eq:squarkmassrelation}) results in  $m_{\tilde{b}_2} < 1000$ GeV, we calculate also the rate for 
\begin{align*}
\label{eq:third-gen-production2}
pp\to\tilde{b}_2\tilde{b}_2^* \quad .
\end{align*}
In such scenarios it can happen that the calculated mass for $\tilde b_2$ is actually smaller
than the input value for $\tilde b_1$. If this happens we flip the resulting mass hierarchy
as we want to keep the default $m_{\tilde b_1}< m_{\tilde b_2}$. We denote those as `flipped'
point/case and treat them separately at the end of this section.

\begin{figure}[t]
\includegraphics{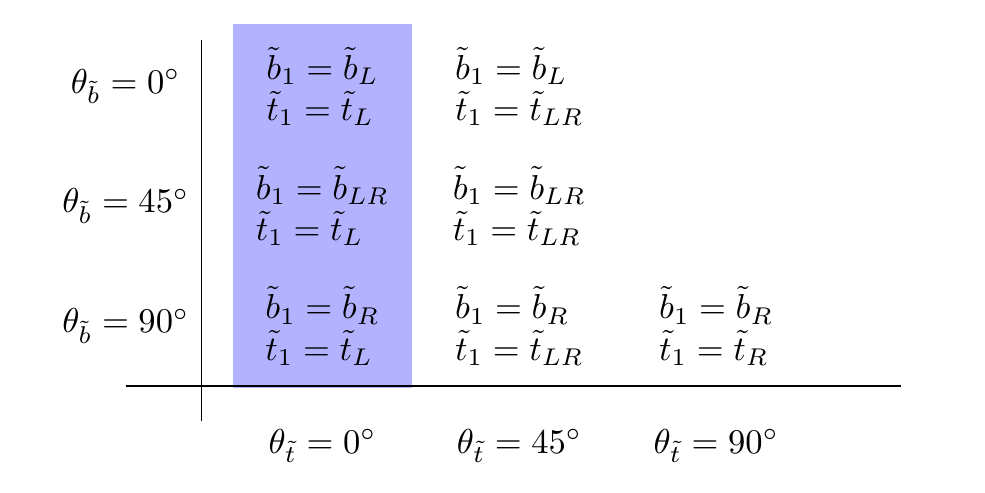}
\caption{Infographic explaining the $3 \times 3$ blocks at the various parameter points
in Figs.~\ref{fig:mSnu60mu110}--\ref{fig:mSnu60mu590}: they correspond to the different
combinations of sbottom and stop mixing angles shown here. The case $\tilde \theta_t$
is special as there eq.~(\ref{eq:squarkmassrelation}) needs to be considered.
}
\label{fig:Expalanation3by3blocks}
\end{figure}

\begin{figure}
\centering
\includegraphics[scale=0.45]{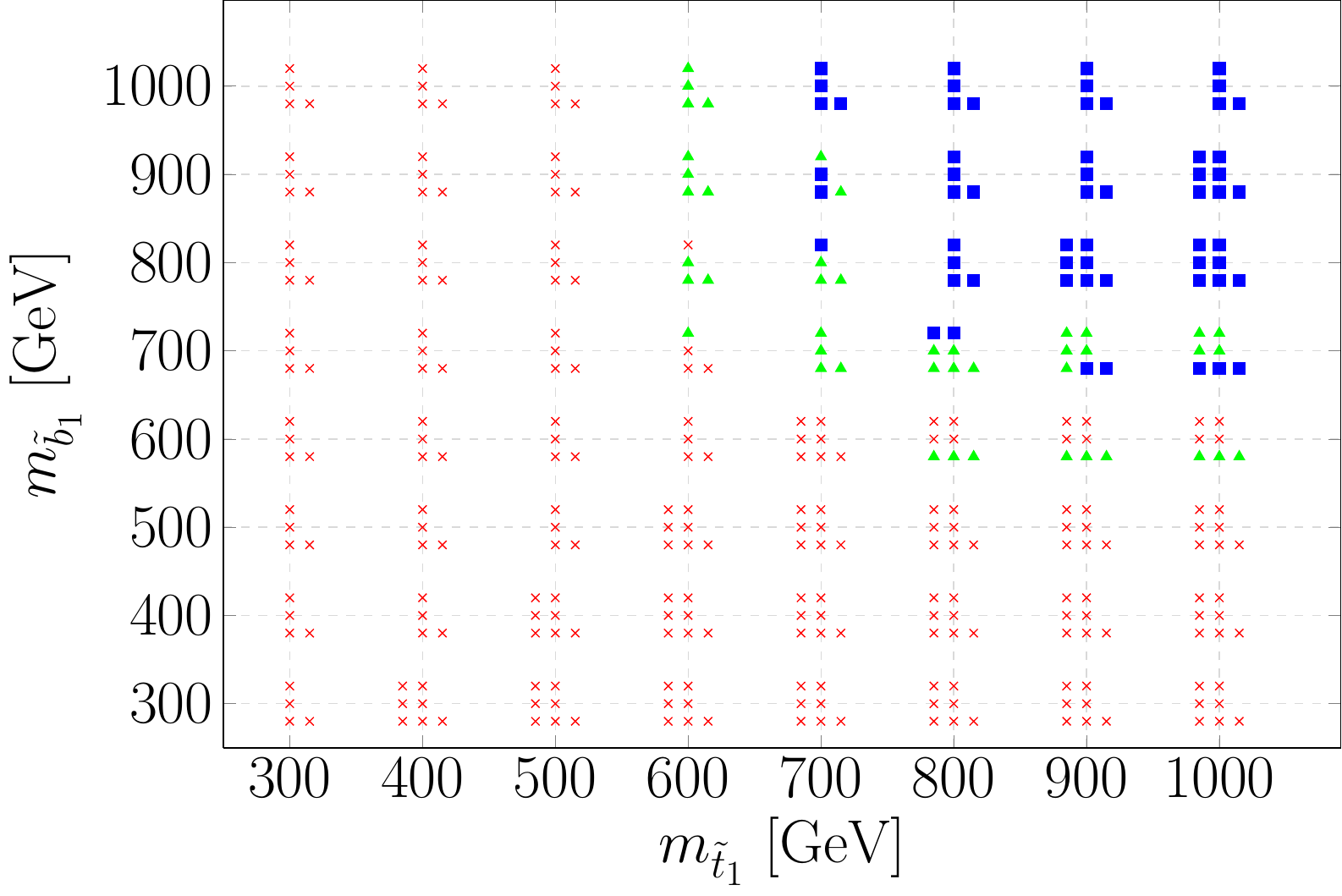} 
\includegraphics[scale=0.45]{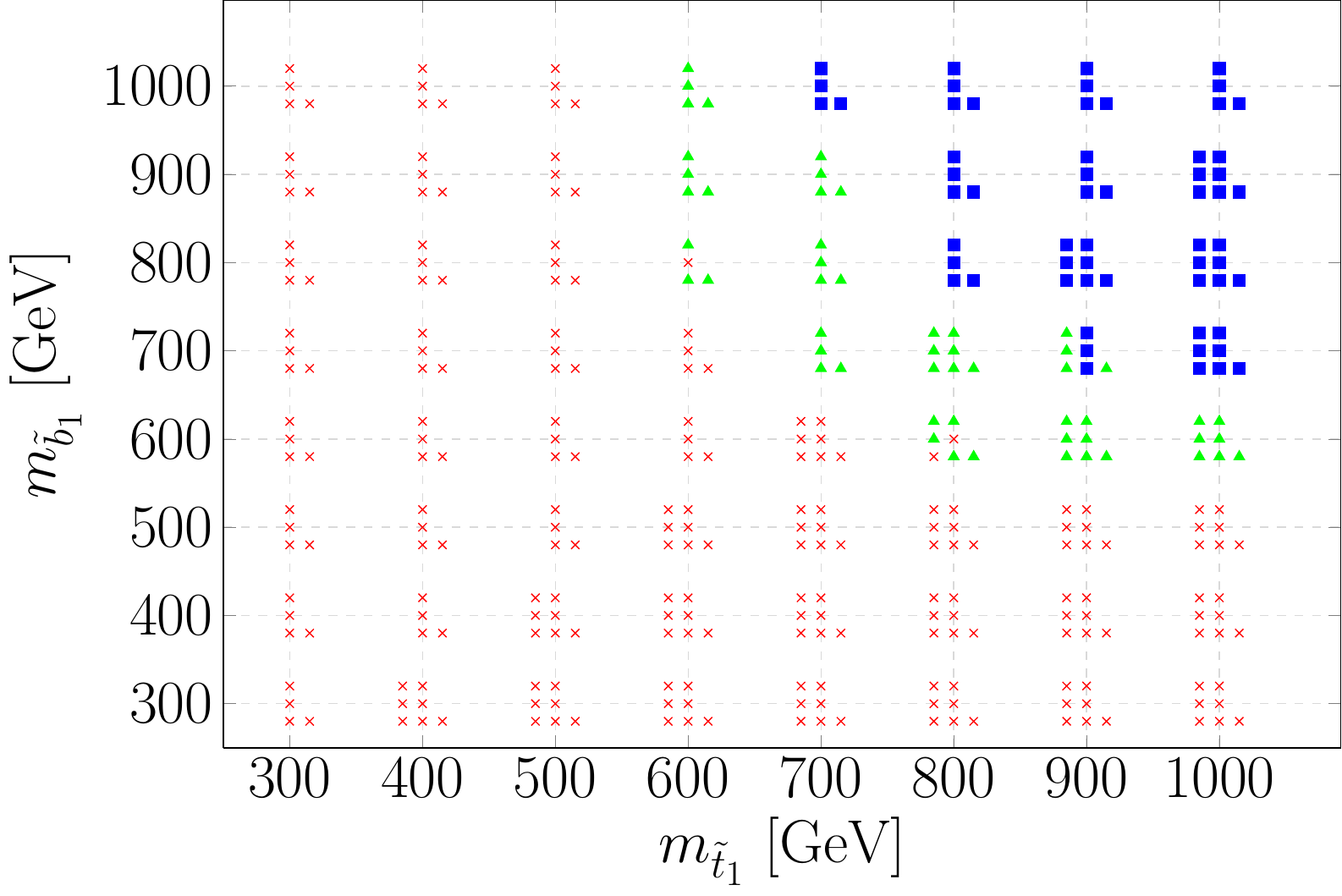} 
\caption{Parameter space points excluded or allowed by the searches listed in Tab.\ 
\ref{tab:ListOfAnalyses} for $\mu=110$ GeV, $m_{\tilde{\nu}_R}=60$ GeV, 
$\tan\beta=10$ (left) and for $\tan\beta=50$ (right). At each 
($m_{\tilde{t}},m_{\tilde{b}_1}$) mass pair one finds a $3\times 3$ block representing the 
different combinations of the mixing angles in the $\tilde{t}$ and the $\tilde{b}$ sector (see Fig. 
\ref{fig:Expalanation3by3blocks}). 
{$\red \times$} designate points excluded by \texttt{CheckMATE}, 
{$\green \blacktriangle$} designate points as ambigous, 
while {$\blue\blacksquare$} indicate points still allowed by the considered ATLAS analyses.
}
\label{fig:mSnu60mu110}
\end{figure}

\begin{figure}[t]
\centering
\includegraphics[scale=0.45]{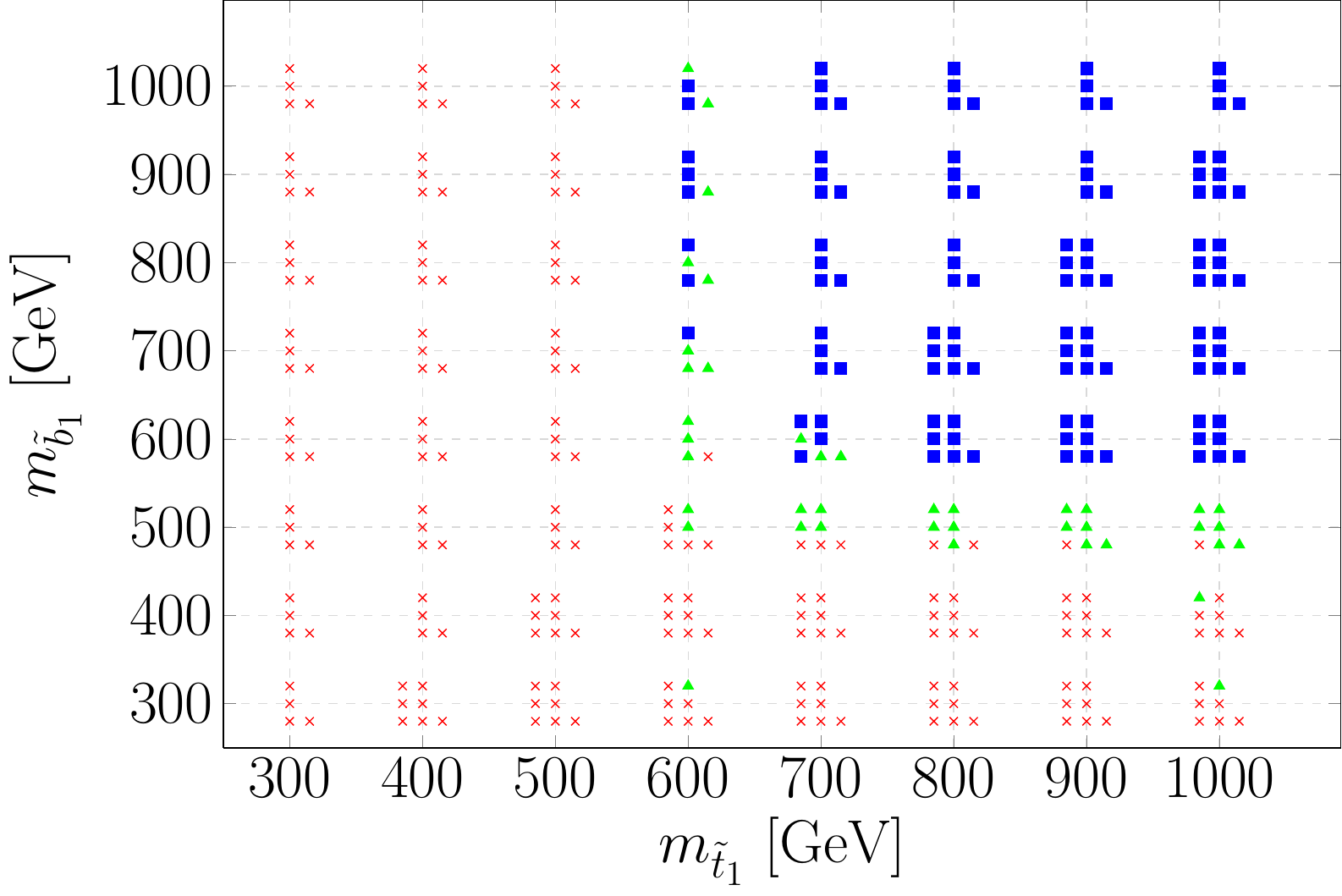}
\includegraphics[scale=0.45]{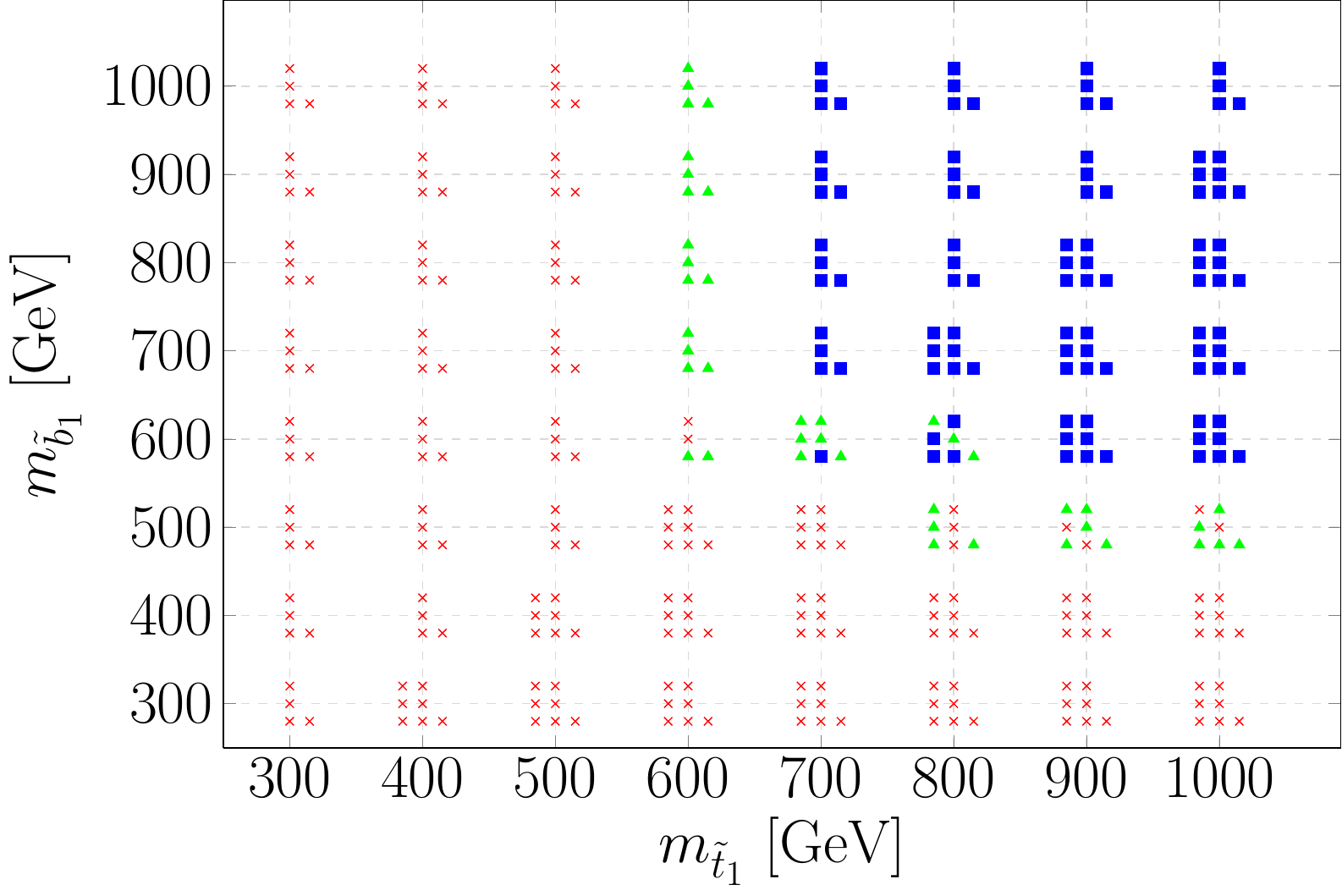}
\caption{The same as Fig.~\ref{fig:mSnu60mu110} but for  $m_{\tilde{\nu}_R}=100$ GeV.}
\label{fig:mSnu100mu110}
\end{figure}

\begin{figure}[t]
\centering
\includegraphics[scale=0.45]{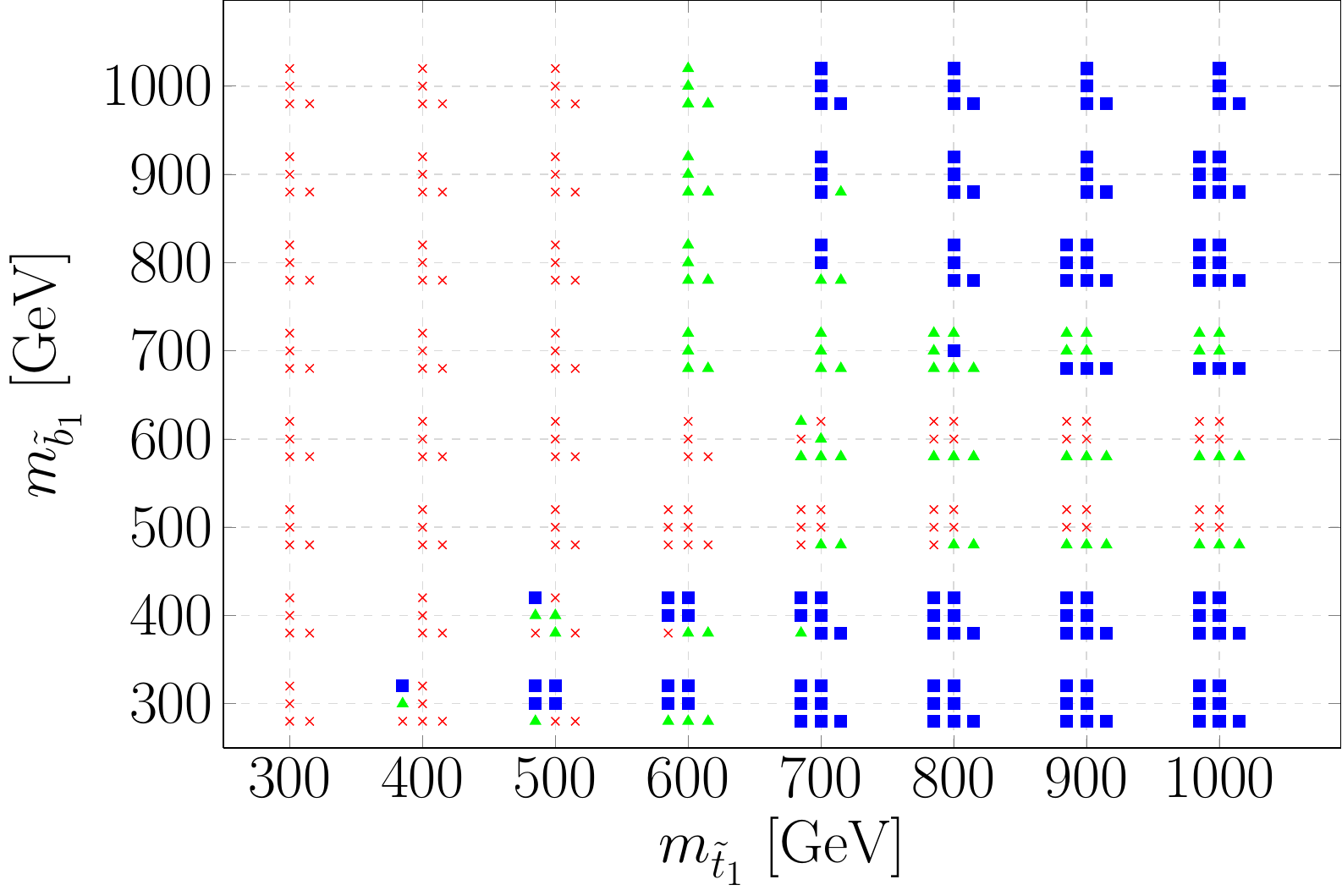} 
\includegraphics[scale=0.45]{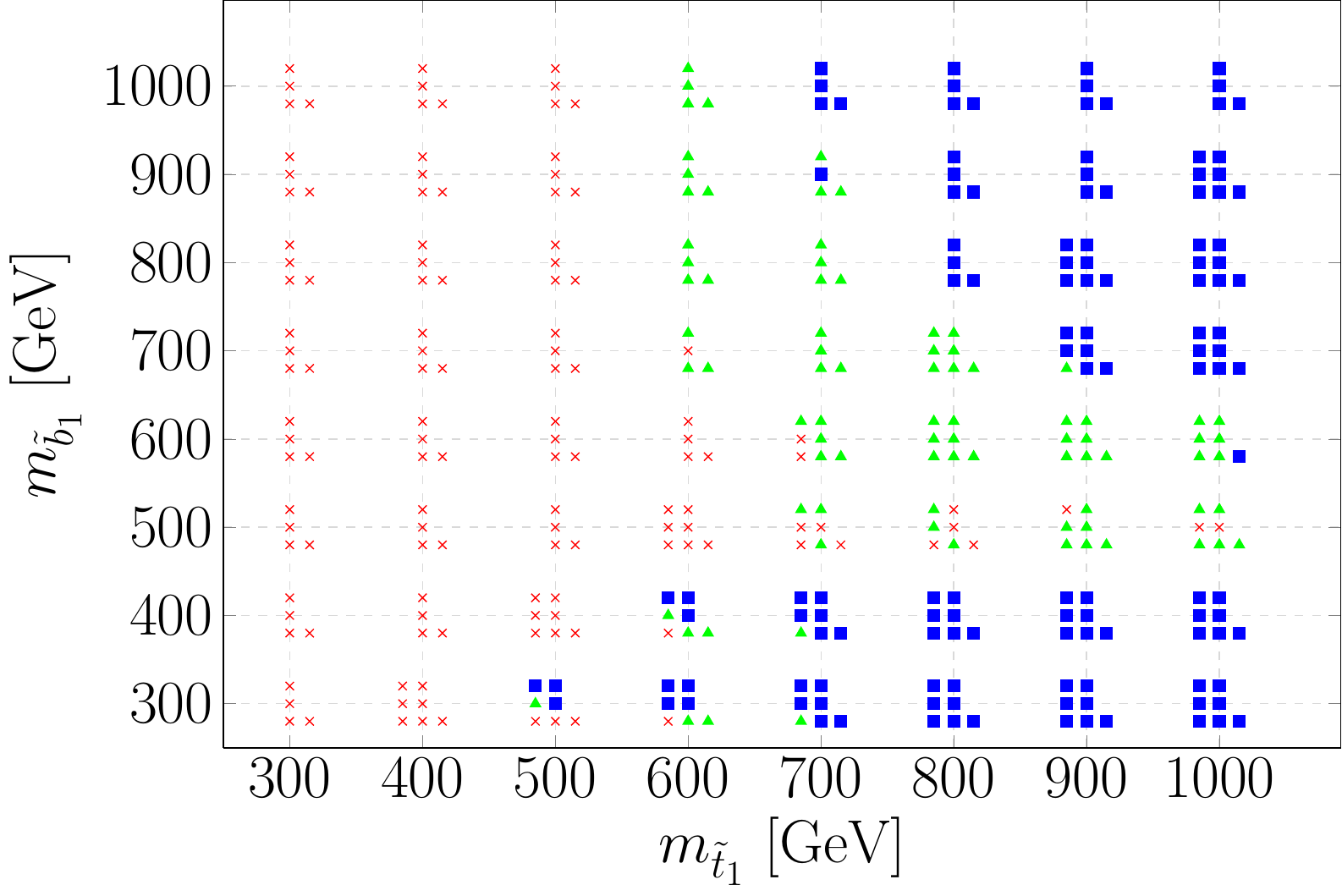} 
\caption{The same as Fig.~\ref{fig:mSnu60mu110} but for  $\mu=290$ GeV and
 $m_{\tilde{\nu}_R}=200$ GeV.}
\label{fig:mSnu200mu290}
\end{figure}

\begin{figure}[t]
\centering
\includegraphics[scale=0.45]{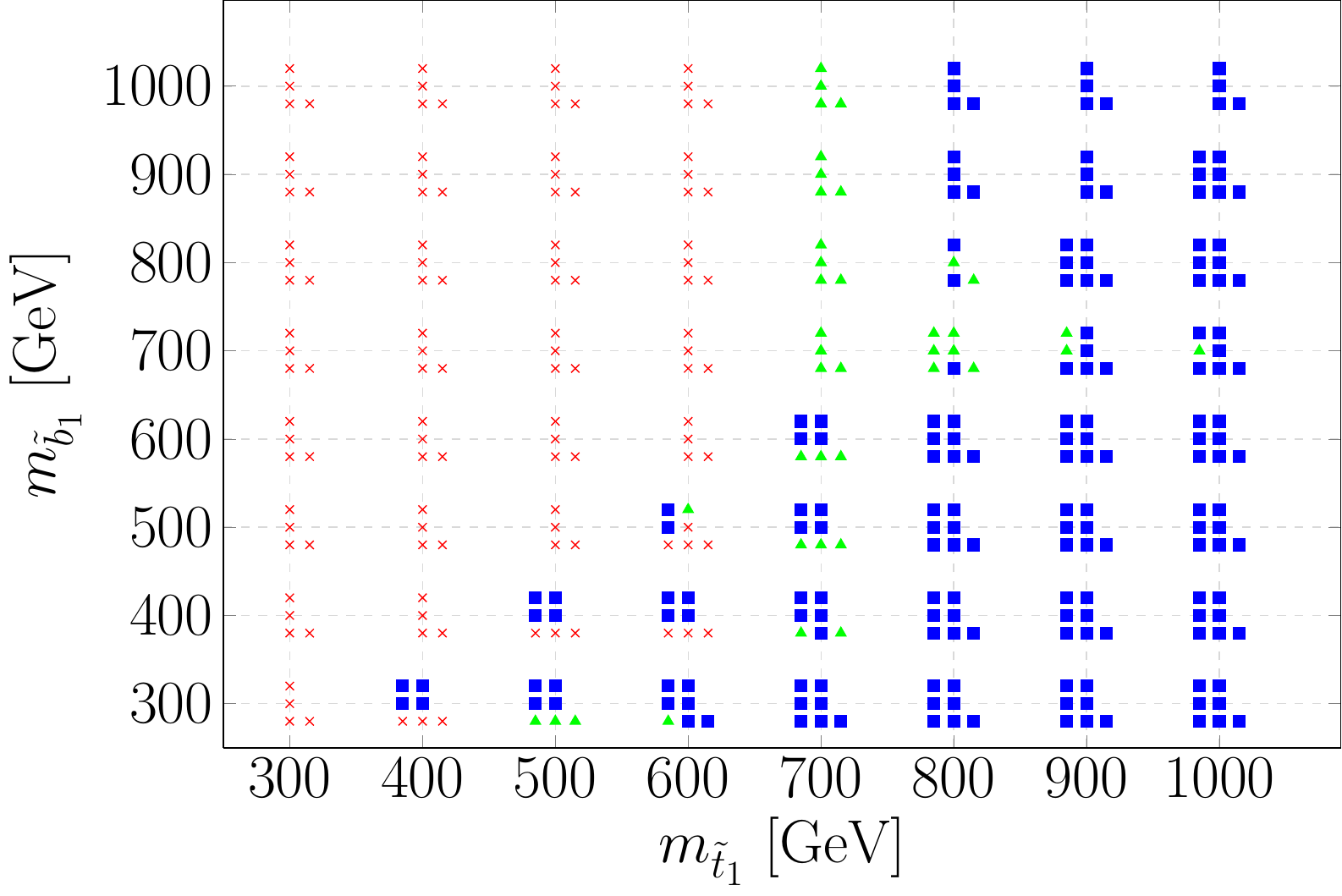} 
\includegraphics[scale=0.45]{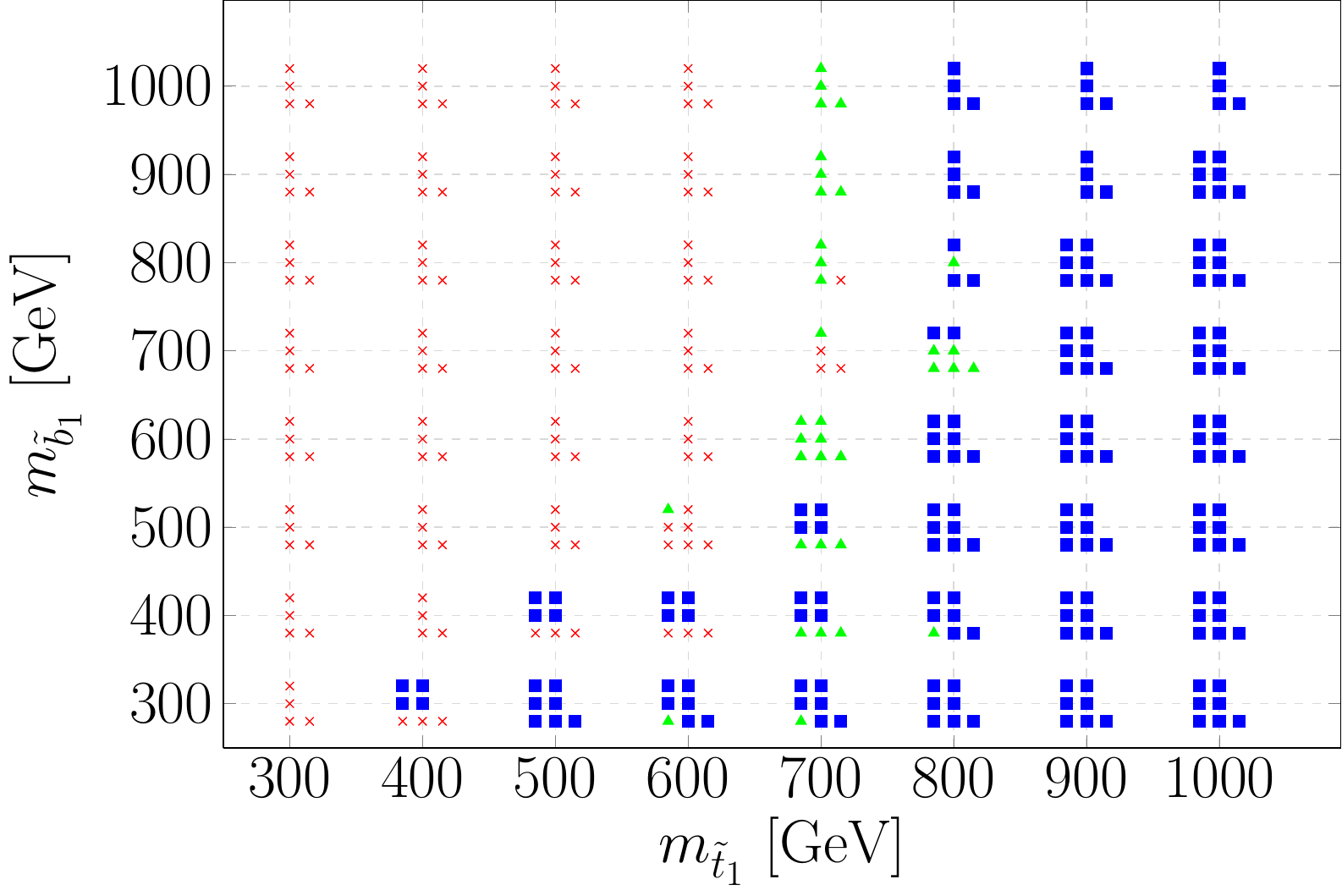} 
\caption{The same as Fig.~\ref{fig:mSnu60mu110} but for  $\mu=490$ GeV and
 $m_{\tilde{\nu}_R}=200$ GeV.}
\label{fig:mSnu200mu490}
\end{figure}

\begin{figure}[t]

\centering
\includegraphics[scale=0.45]{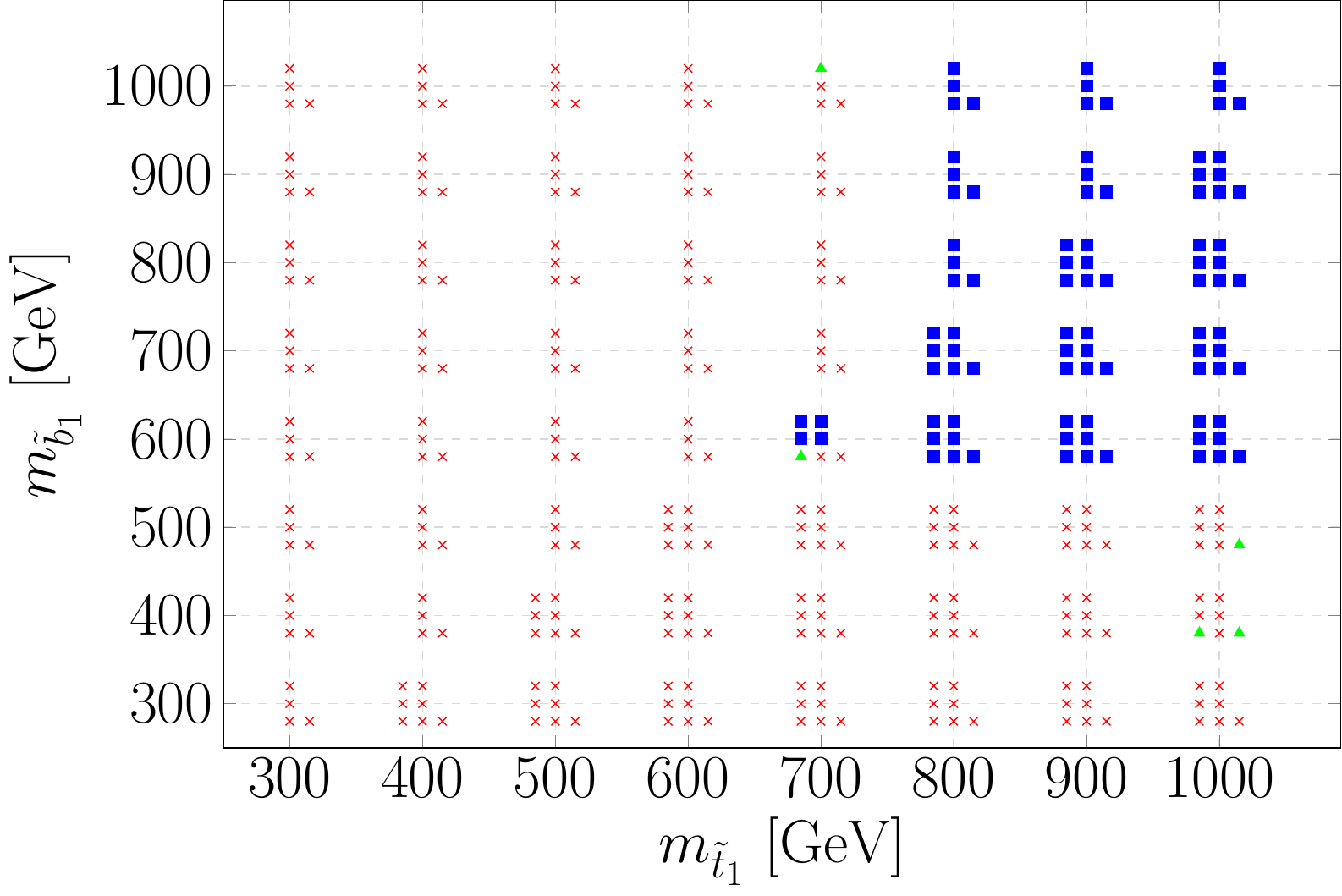} 
\includegraphics[scale=0.45]{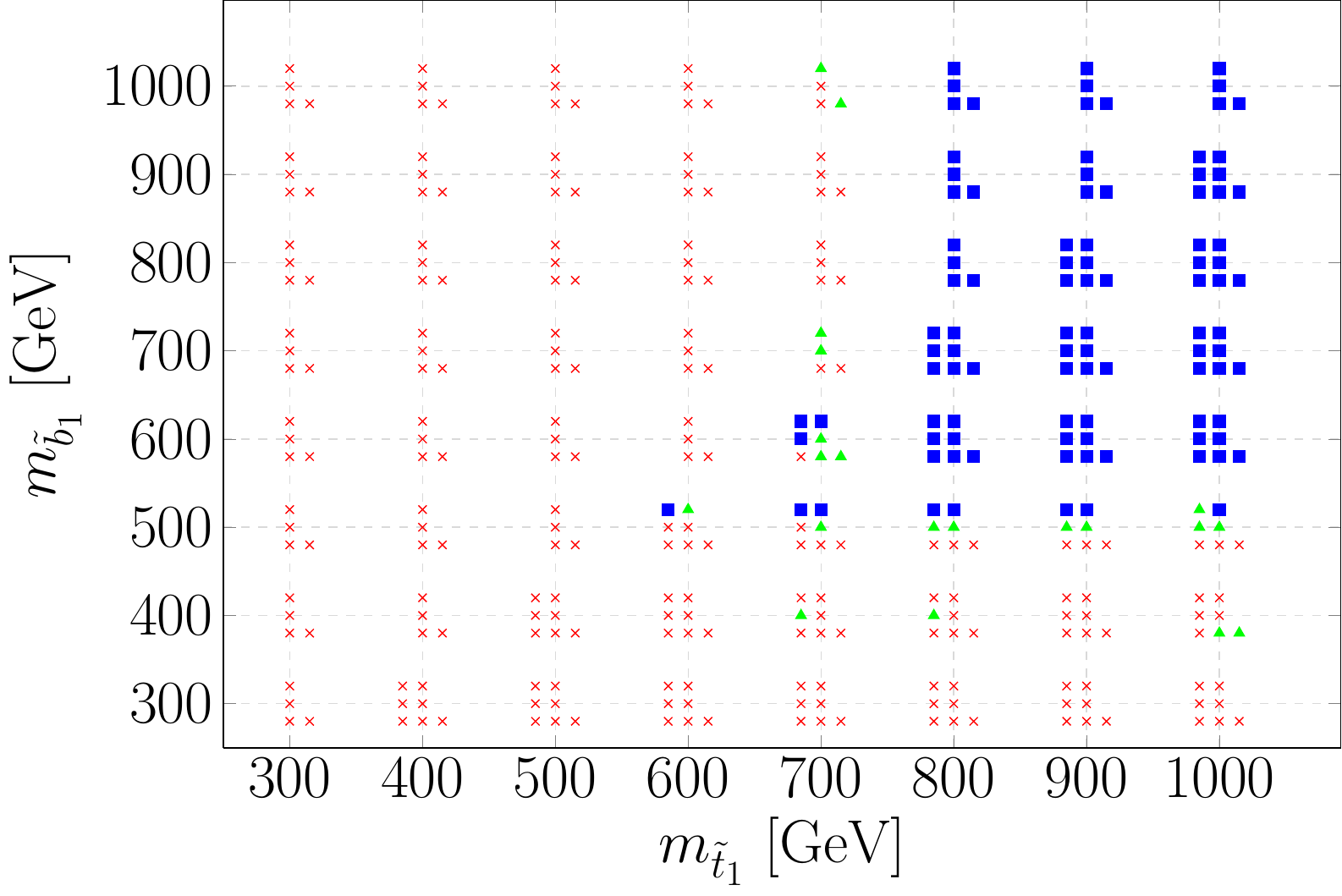} 
\caption{The same as Fig.~\ref{fig:mSnu60mu110} but for  $\mu=590$ GeV.}
\label{fig:mSnu60mu590}
\end{figure}

We start by presenting our findings for various parameter combinations in the
$m_{\tilde t_1}$-$m_{\tilde b_1}$ and refer for 
 more details to the coming ref.~\cite{DissLukas}. 
A short summary explaining the entries of the plots in 
Figs.~\ref{fig:mSnu60mu110}--\ref{fig:mSnu60mu590} is given in 
Fig.~\ref{fig:Expalanation3by3blocks}. At every point in the $m_{\tilde t_1}$-$m_{\tilde b_1}$
we display all considered combinations of mixing angles in the stop and sbottom sector by 
slightly shifting the result depending on the respective mixing angle:  The right column represents 
scenarios with a right stop and a right sbottom, i.e.\ $\tilde{t}_1 = \tilde{t}_R$.  The middle 
column belongs to a maximally mixed stop whereas the left column  consists of all points with a 
left stop $\tilde{t}_1 = \tilde{t}_L$, i.e. $\theta_{\tilde{t}}=0$. In the scenarios of 
this column we use the tree-level relation (\ref{eq:squarkmassrelation}) as discussed above.
Note, that even if no flipping of the sbottom masses takes place, one might have additional
signals coming from the production of the heavier sbottom.
The different rows belong to different mixing angles in the sbottom sector:
The top one corresponds to  $\theta_{\tilde{b}}=0$, the middle one to 
$\theta_{\tilde{b}}=45^{\circ}$ and the lowest one to $\theta_{\tilde{b}}=90^{\circ}$. 

In Figures~\ref{fig:mSnu60mu110}--\ref{fig:mSnu60mu590} we display 
those points where the mass ordering of the sbottoms is not changed. We  comment on
the flipped sbottom mass  scenarios at the end of this section as they hardly
lead to new features. 
Figure \ref{fig:mSnu60mu110} shows the case with a rather hierarchical mass 
spectrum. As a result the final states contain several hard leptons implying
the stop and sbottom masses below about 500 GeV can be excluded. For larger values
the cross sections get so much reduced that one would have to take higher order
corrections into account. Moreover, these cases would have required more detailed
experimental investigations as the experimental cuts are most likely not
optimized for these cases. In case that both, ${\tilde{b}_1}$ and ${\tilde{t}_1}$,
have masses above about 800 GeV, the production cross section gets too small
and, thus, these scenarios are strictly allowed. Comparing the two values of
$\tan\beta$ we find only a small effect: the bounds are somewhat softer for large
$\tan\beta$ due to the smaller number of produced top-quarks. Note, the composition
of the lepton flavours does not depend on $\tan\beta$ here as we have a pure
right-handed sneutrino. In case that also the left-handed sneutrino mass parameter
gets smaller, we would expect a further reduction of the bounds as the number of
$\tau$-leptons would increase with a simultaneous reduction of electrons and muons in 
the final state.

In Figure \ref{fig:mSnu100mu110} we show a scenario with a small mass difference
$m_{\tilde{\chi}^\pm}-m_{\tilde{\nu}_R} = 10$~GeV 
leading to softer leptons. As a consequence the bounds on the stop and sbottom masses
get weakened by at least 100 GeV. A further reduction happens if also the mass difference
between the squarks and the sneutrino gets small because then also the resulting
$b$-quarks are too soft to be detected. An example is show in Fig.~\ref{fig:mSnu200mu290}
where we have set $\mu=290$~GeV and $m_ {\tilde \nu_R}=200$~GeV. In particular
the sbottom searches in the $b\slashed{E}_T$ become less important because of the
requirement that the jet has to have a $p_T$ of at least 150 GeV in the analysis
 \texttt{atlas\_1308\_2631} \cite{Aad:2013ija} which is not met in case of light sbottoms
 with a mass of 300-400 GeV. The situation is different for stops with such a mass as
there the decay into $b \tilde \chi^+_1 \to b l \tilde \nu_R$ is present and thus the
lepton searches are effective. 

An increase in $\mu$ leads to a merger of the allowed region created by the compressed mass 
spectrum with the high-masses allowed region, as can be seen in the trend going from 
Fig.~\ref{fig:mSnu200mu290} to Fig.~\ref{fig:mSnu200mu490}.
However the size of this allowed region is still dependent on the mass of  $\tilde{\nu}_R$  
as can be observed in a comparison of Fig.~\ref{fig:mSnu200mu490} and Fig.~\ref{fig:mSnu60mu590}.
In particular in the later case only three-body decays of the squarks are possible if their
masses are below 600 GeV resulting in hard $b$-jets and leptons and thus to a larger
excluded region in the $m_{\tilde{t}_1}$--$m_{\tilde{b}_1}$ plane compared to the previous case.

\begin{figure}[t]
\centering
\begin{tabular}{ccccc}
a) & & &b) & \\
c) &\includegraphics[scale=0.43]{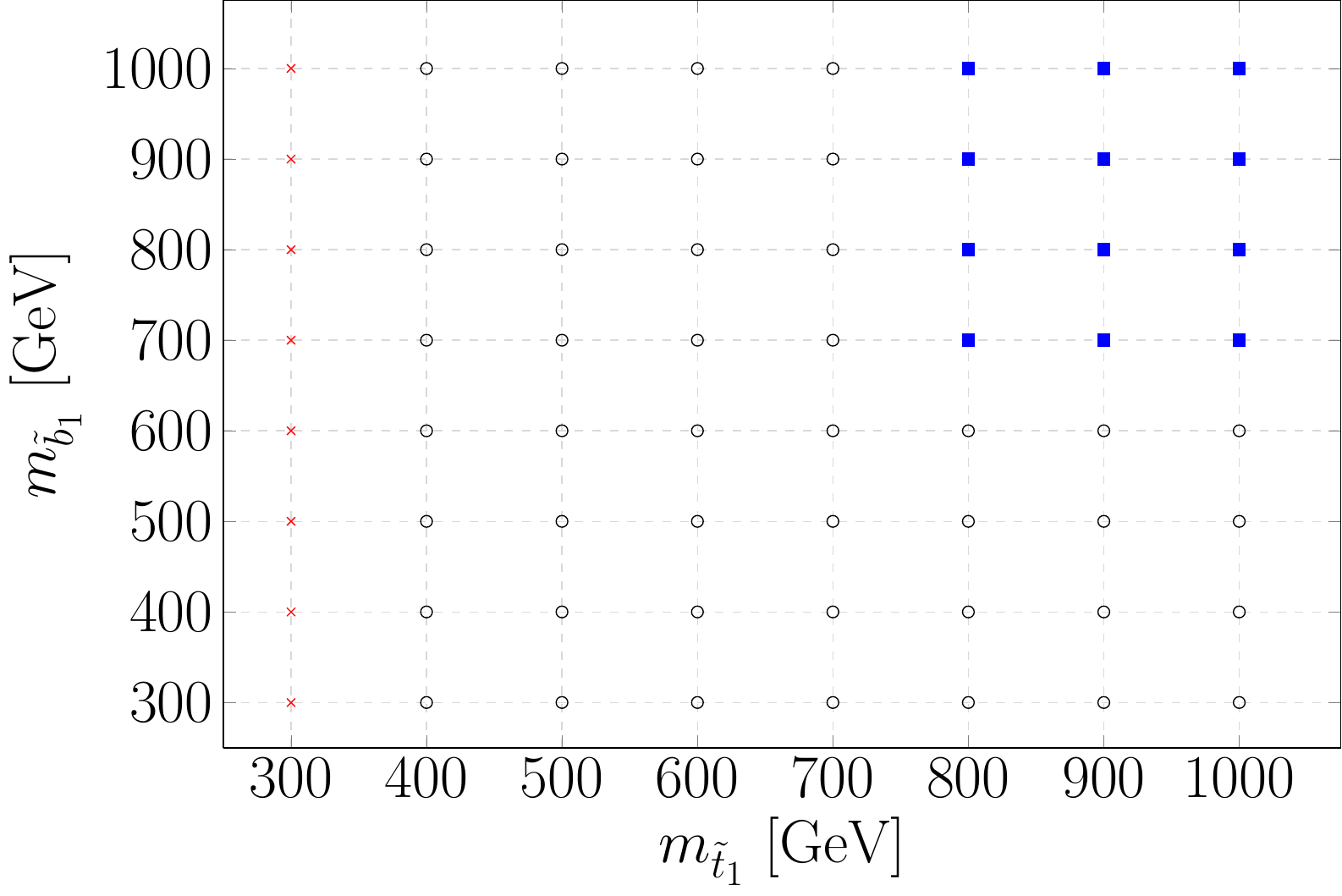} &
&d) &\includegraphics[scale=0.43]{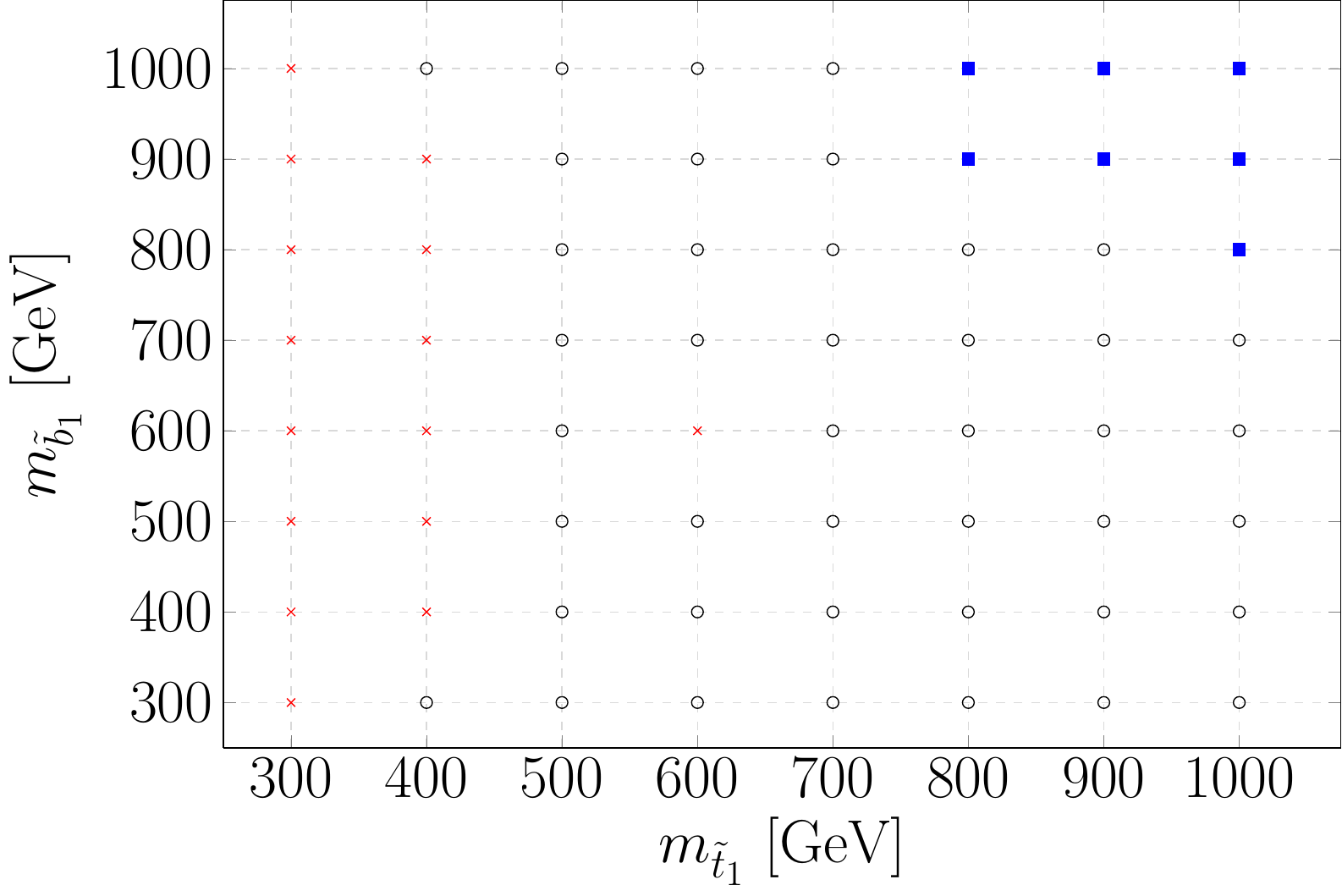} \\
 & \includegraphics[scale=0.43]{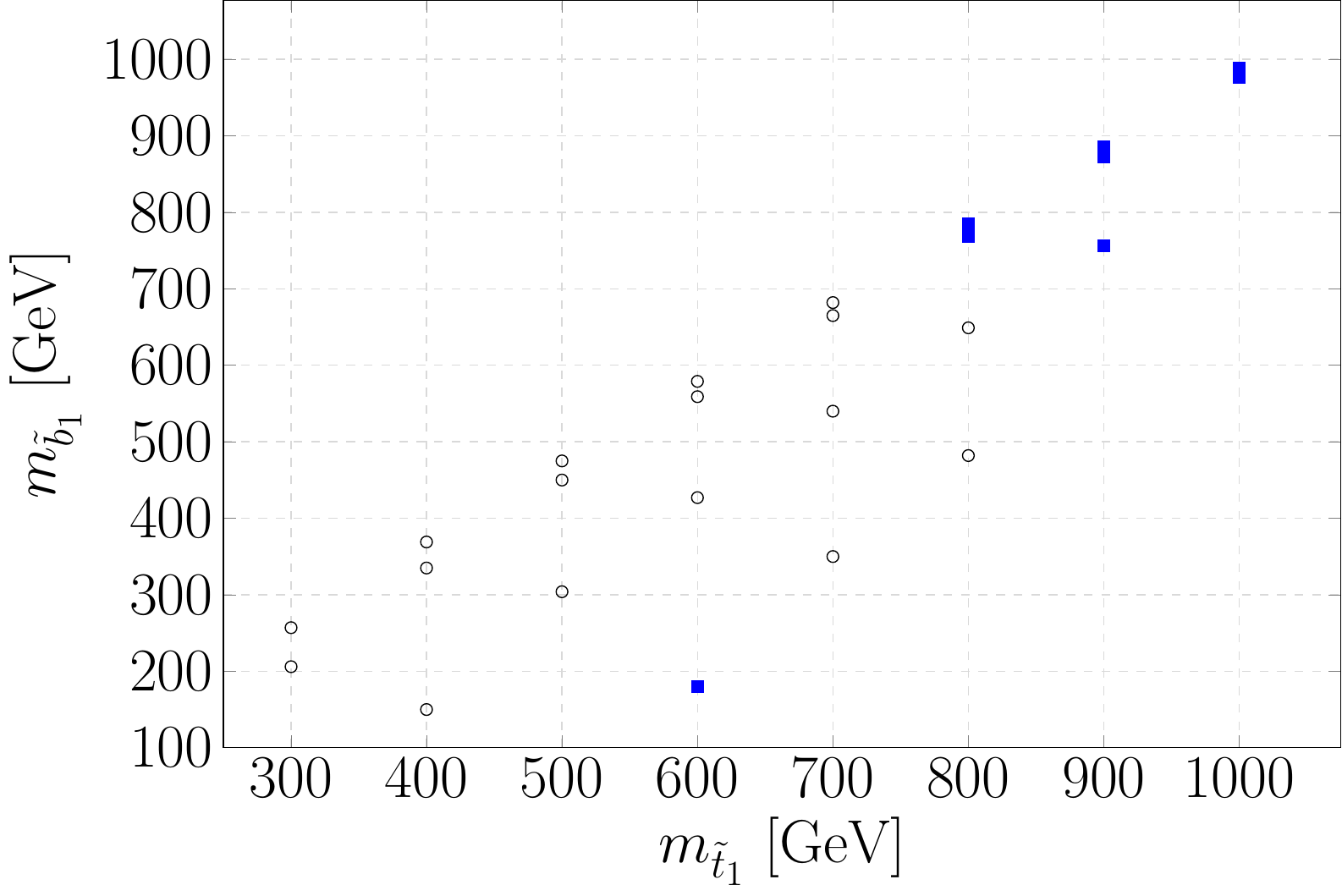} &
& & \includegraphics[scale=0.43]{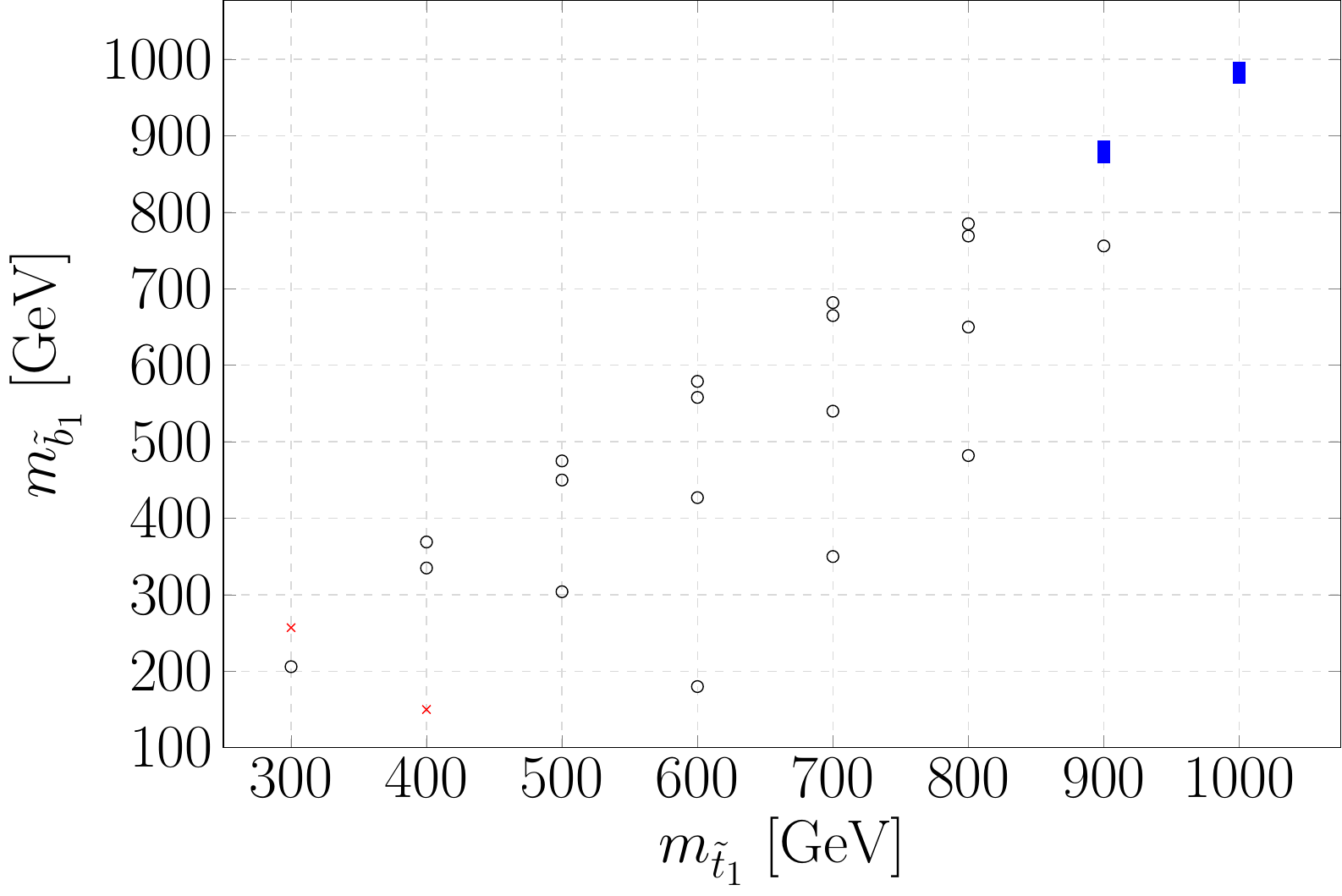}  
\end{tabular}
\caption{Summary plots in the $m_{\tilde{t}_1}$--$m_{\tilde{b}_1}$ plane
marginalized over all allowed and considered $\tilde \theta_b$,
$\tilde \theta_t$, $\tan\beta$, $\mu$ and $m_{\tilde \nu_R}$
combinations. A point denoted by {$\red \times$} is excluded for all combinations, 
for a point with $\circ$ the  exclusion depends on the  parameter space combination, 
while points with a {$\blue\blacksquare$} are allowed for all parameter combinations.
The different plots correspond to: a) Counting 'ambiguous' points conservatively as 
allowed in cases where no flipping in the 
sbottom mass hierarchy occurred. b) Counting 'ambiguous' points as excluded 
in cases where no flipping in the 
sbottom mass hierarchy occurred. c) Counting 'ambiguous' points conservatively as 
allowed in cases where a flipping in the 
sbottom mass hierarchy occurred. d) Counting 'ambiguous' points as excluded 
in cases where a flipping in the 
sbottom mass hierarchy occurred. 
}
\label{fig:condensed-exclusion}
\end{figure}

In Figure~\ref{fig:condensed-exclusion} we summarize the exclusion statements in a condensed
way. We put the 'ambiguous' either conservatively to the set of allowed points (left column)
or to the excluded set (right column) to get a rough idea of the underlying uncertainty.
The upper row gives the cases with an unflipped sbottom mass spectrum. We see in plot a)
that only the case of $m_{\tilde t_1} =300$ GeV is strictly excluded if the ambiguous  points are
counted as allowed. The corner with $m_{\tilde{b}_1} \geq 700$GeV and 
$m_{\tilde{t}_1} \geq 800$~GeV is not constrained at all which was expected due to the
small production cross section.
The exclusion statement for all other points is strongly dependent on the underlying parameter combinations as can be seen from the discussion of the individual cases before.
A comparison with the right column shows that the range of excluded stop masses for
all other parameter combinations gets somewhat larger if the ambiguous points are counted
as excluded indicating that this region deserves further investigations. This also correct  for
some parameter combinations in the 
the higher stop and sbottom mass region close to 800 GeV. A particular case
is $m_{\tilde t_1} = m_{\tilde b_1} = 600$~GeV which seems now to be excluded
for all parameter combinations. However, this has to be taken with a grain of salt
as we have restricted $\mu \le 590$~GeV and, thus, having always two-body decays.
We have also checked that this point is not excluded, which means having
$r^c_{obs}$ values below 2/3, for some parameter combinations
if only three body decays of the squarks are allowed  by setting $\mu=610$~GeV for this
particular case.

The lower row gives the cases where the sbottom mass ordering is flipped. Here
we exclude right from the start all points where the sbottom is lighter than the sneutrino. 
 We see
that these cases are consistent with the findings in the upper row. However, in
addition we have here a couple of points where the lighter sbottom has a mass below
300 GeV. It is intriguing that in the rather low mass region the exclusion depends
on the parameter combinations in the conservative case (left side). 
The reason for this lies in the particular mass hierarchies: in cases, where the decay $\tilde{t}_1\to W^+ \tilde{b}_1$ is the only viable one due to a large $\mu$, the value of the sneutrino mass 
determines the $p_T$ and visibility of the exiting $b$ quark which helps to pass the particular analysis' jet requirements in 
addition to the jets coming from the $W$. Moreover, the mass splitting between $\tilde{t}_1$ and $\tilde{b}_1$
 is rather small leaving little phase space for the $W$-boson which in turn implies that the resulting  jet energies hardly pass the required $p_T$-cuts.

A comparison of our exclusion with those ref.~\cite{Guo:2013asa} shows that due the fact, that
we do not only consider $\cos\theta_{\tilde t}=0$, implying $\tilde t_1=\tilde t_R$, but 
consider three different values,
we do not reach their strict exclusions of $m_{\tilde{t}_1}$ up to 900 GeV. However,
if for a pure $\tilde t_R$ we do not reach this value which is due to the differences
between \texttt{CheckMATE} and \texttt{PGS}.  Similarly, we get lower bounds on
the mass bounds of the chargino and sneutrinos from the direct chargino production:
we get roughly a bound of $m_{\tilde{\chi}^\pm} \sim 300$ GeV and $m_{\tilde \nu_R} \sim 120 $ GeV
whereas in \cite{Guo:2013asa} the bounds $m_{\tilde{\chi}^\pm} \sim 550$ GeV and 
$m_{\tilde \nu_R} \sim 300 $ GeV. The main difference is the more conservative
$r^c_{obs}$-measure of \texttt{CheckMATE} given in eq.~(\ref{eq:r-value-definition}) whereas
in \texttt{PGS} the measure $R=\frac{N_{NP}}{N_{limit}}$ is used with $N_{NP}$ and $N_{limit}$
 being the number of events from the new physics and the experimental upper bound, respectively.

\section{Conclusions}
\label{sec:conclusion}
In this paper we have studied a variant of the natural supersymmetric models containing
a right-handed sneutrino LSP as it can 
emerge as effective model from models with extended gauge groups.
We have considered different scenarios taking also into account the case that at least
one of the sbottoms gets light. The latter can in particular occur if the lighter stop
is essentially left-handed. Provided that the mass difference is not too small we
find that only stop masses of up to about 300 GeV can be excluded in this scenario
independent of its nature. Depending on the scenario, where the main parameters are the
stop mixing angles and the ratio of stop mass to $\mu$ to sneutrino mass, one can
exclude larger masses of up to 700 GeV. Here we roughly confirm the findings of 
\cite{Guo:2013asa} where the case of a pure $\tilde t_R$ had been considered.
In addition we obtain similar results for the sbottoms. Last but not least we find that
the bounds are somewhat weaker in case of large $\tan\beta$ as the number of produced
$t$-quarks from the squark decays is smaller which in turn reduces the average 
number of hard leptons in the final state.

\section*{Acknowledgements}
We thank D.~Schmeier and J.~Tattersall for several very helpful comments and  annotations
to {\tt CheckMATE} as well as J.~Reuter, T.~Ohl and W.~Kilian in case of {\tt WHIZARD}.
This work  has been supported by  the DFG, project nr.\ PO-1337/3-1.


\end{document}